\documentclass[12pt,letterpaper]{article}
\usepackage{amssymb}
\usepackage{amsmath}
\allowdisplaybreaks
\usepackage{amscd}
\usepackage{graphicx}
\usepackage{tabularx,booktabs}
\usepackage{array}
\newcolumntype{Z}{>{\centering\let\newline\\\arraybackslash\hspace{0pt}}X}
\usepackage[footnotesize,bf,textfont={it},margin=1 cm]{caption}%sets captions titles to bold, and the whole caption to footnotesize

\def\be{\begin{equation}}
\def\ee{\end{equation}}
\def\frc#1#2{\relax\ifmmode{\textstyle\frac{#1}{#2}} % A small fraction
                    \else$\frac{#1}{#2}$\fi}         % good in text

\def\hj{{\hat\jmath}}
\def\hk{{\hat k}}
\def\hi{{\hat\imath}}

\def\bj#1{{}_{\rm #1}}   
\def\fracm#1#2{\hbox{\large{${\frac{{#1}}{{#2}}}$}}}

\def\pp{{\mathchoice
          {%displaystyle
              \kern 1pt%
              \raise 1pt
              \vbox{\hrule width5pt height0.4pt depth0pt
                    \kern -2pt
                    \hbox{\kern 2.3pt
                          \vrule width0.4pt height6pt depth0pt
                          }
                    \kern -2pt
                    \hrule width5pt height0.4pt depth0pt}%
                    \kern 1pt
           }
            {%textstyle
              \kern 1pt%
              \raise 1pt
              \vbox{\hrule width4.3pt height0.4pt depth0pt
                    \kern -1.8pt
                    \hbox{\kern 1.95pt
                          \vrule width0.4pt height5.4pt depth0pt
                          }
                    \kern -1.8pt
                    \hrule width4.3pt height0.4pt depth0pt}%
                    \kern 1pt
            }
            {%scriptstyle
              \kern 0.5pt%
              \raise 1pt
              \vbox{\hrule width4.0pt height0.3pt depth0pt
                    \kern -1.9pt  %[e]=0.15pt
                    \hbox{\kern 1.85pt
                          \vrule width0.3pt height5.7pt depth0pt
                          }
                    \kern -1.9pt
                    \hrule width4.0pt height0.3pt depth0pt}%
                    \kern 0.5pt
            }
            {%scriptscriptstyle
              \kern 0.5pt%
              \raise 1pt
              \vbox{\hrule width3.6pt height0.3pt depth0pt
                    \kern -1.5pt
                    \hbox{\kern 1.65pt
                          \vrule width0.3pt height4.5pt depth0pt
                          }
                    \kern -1.5pt
                    \hrule width3.6pt height0.3pt depth0pt}%
                    \kern 0.5pt%}
            }
        }}

\def\mm{{\mathchoice
                       {%displaystyle
                             \kern 1pt
               \raise 1pt    \vbox{\hrule width5pt height0.4pt depth0pt
                                  \kern 2pt
                                  \hrule width5pt height0.4pt depth0pt}
                             \kern 1pt}
                       {%textstyle
                            \kern 1pt
               \raise 1pt \vbox{\hrule width4.3pt height0.4pt depth0pt
                                  \kern 1.8pt
                                  \hrule width4.3pt height0.4pt depth0pt}
                             \kern 1pt}
                       {%scriptstyle
                            \kern 0.5pt
               \raise 1pt
                            \vbox{\hrule width4.0pt height0.3pt depth0pt
                                  \kern 1.9pt
                                  \hrule width4.0pt height0.3pt depth0pt}
                            \kern 1pt}
                       {%scriptscriptstyle
                           \kern 0.5pt
             \raise 1pt  \vbox{\hrule width3.6pt height0.3pt depth0pt
                                  \kern 1.5pt
                                  \hrule width3.6pt height0.3pt depth0pt}
                           \kern 0.5pt}
                       }}

\def\ad{{\kern0.5pt
                   \alpha \kern-5.05pt \raise5.8pt\hbox{$\textstyle.$}\kern
0.5pt}}

\def\bd{{\kern0.5pt
                   \beta \kern-5.05pt \raise5.8pt\hbox{$\textstyle.$}\kern
0.5pt}}

\def\qd{{\kern0.5pt
                   q \kern-5.05pt \raise5.8pt\hbox{$\textstyle.$}\kern
0.5pt}}
\def\Dot#1{{\kern0.5pt
     {#1} \kern-5.05pt \raise5.8pt\hbox{$\textstyle.$}\kern
0.5pt}}

\catcode`@=11
\def\un#1{\relax\ifmmode\@@underline#1\else
        $\@@underline{\hbox{#1}}$\relax\fi}
\catcode`@=12

\def\a{\alpha}
\def\b{\beta}

\def\d{\delta}
\def\e{\epsilon}

\def\g{\gamma}

\def\l{\lambda}
\def\m{\mu}
\def\n{\nu}

\def\r{\rho}
\def\s{\sigma}
\def\t{\tau}

\def\D{\Delta}

\def\O{\Omega}

\def\S{\Sigma}
\def\U{\Upsilon}

\def\dslash{\not{\hbox{\kern-2pt $\partial$}}}
\def\Dslash{\not{\hbox{\kern-4pt $D$}}}
\def\pslash{\not{\hbox{\kern-2.3pt $p$}}}
 \newtoks\slashfraction
 \slashfraction={.13}
 \def\slash#1{\setbox0\hbox{$ #1 $}
 \setbox0\hbox to \the\slashfraction\wd0{\hss \box0}/\box0 }
 
\font\ro=cmsy10                          % font with rope
\def\kcr{{\hbox{\ro \char'170}}}                % right-handed rope
\def\ktl{{\hbox{\ro \char'170}}}        % top end for left-handed rope
\def\ktr{{\hbox{\ro \char'170}}}        % " right
\def\kbl{{\hbox{\ro \char'170}}}        % " bottom left
\def\kbr{{\hbox{\ro \char'170}}}        % " right
\def\plpl{\raise-2pt\hbox{$\raise3pt\hbox{$_+$}\hskip-6.67pt\raise0.0pt
\hbox{$^+$}\hskip 0.01pt$}}
\def\mimi{\raise-2pt\hbox{$\raise3pt\hbox{$_-$}\hskip-6.67pt\raise0.0pt
\hbox{$^-$}\hskip 0.01pt$}} 

\def\bo{{\raise.15ex\hbox{\large$\Box$}}}               % D'Alembertian
\def\pa{\partial}                                       % curly d
\def\TH{{\raise.2ex\hbox{$\displaystyle \bigodot$}\mskip-4.7mu \llap H \;}}
\def\face{{\raise.2ex\hbox{$\displaystyle \bigodot$}\mskip-2.2mu \llap {$\ddot
        \smile$}}}                                      % happy face
\def\Tilde#1{\widetilde{#1}}                    % big tilde
\def\Hat#1{\widehat{#1}}                        % big hat
\def\Bar#1{\overline{#1}}                       % big bar
\def\leftrightarrowfill{$\mathsurround=0pt \mathord\leftarrow \mkern-6mu
        \cleaders\hbox{$\mkern-2mu \mathord- \mkern-2mu$}\hfill
        \mkern-6mu \mathord\rightarrow$}
\def\dvec#1{\vbox{\ialign{##\crcr
        \leftrightarrowfill\crcr\noalign{\kern-1pt\nointerlineskip}
        $\hfil\displaystyle{#1}\hfil$\crcr}}}           % <--> accent
\def\dt#1{{\buildrel {\hbox{\LARGE .}} \over {#1}}}     % dot-over for sp/sb
\def\fracm#1#2{\hbox{\large{${\frac{{#1}}{{#2}}}$}}}
\def\frac#1#2{{\textstyle{#1\over\vphantom2\smash{\raise.20ex
        \hbox{$\scriptstyle{#2}$}}}}}                   % fraction
\def\sfrac#1#2{{\vphantom1\smash{\lower.5ex\hbox{\small$#1$}}\over
        \vphantom1\smash{\raise.4ex\hbox{\small$#2$}}}} % alternate fraction
\def\bfrac#1#2{{\vphantom1\smash{\lower.5ex\hbox{$#1$}}\over
        \vphantom1\smash{\raise.3ex\hbox{$#2$}}}}       % "
\def\afrac#1#2{{\vphantom1\smash{\lower.5ex\hbox{$#1$}}\over#2}}    % "
\def\on#1#2{\mathop{\null#2}\limits^{#1}}               % arbitrary accent
\def\pa{\partial}      
\newcommand{\bm}[1]{\mbox{\boldmath$#1$}}

\def\ad{{\dot{\alpha}}}
\def\bd{{\dot{\beta}}}

\font\ro=cmsy10                          % font with rope
\def\kcr{{\hbox{\ro \char'170}}}                % right-handed rope
\def\ktl{{\hbox{\ro \char'170}}}        % top end for left-handed rope
\def\ktr{{\hbox{\ro \char'170}}}        % " right
\def\kbl{{\hbox{\ro \char'170}}}        % " bottom left
\def\kbr{{\hbox{\ro \char'170}}}        % " right
\parskip=\medskipamount                 % space between paragraphs (LaTeX)
\thicklines                         % thick straight lines for pictures (LaTeX)

\def\border{                                            % border
        \setlength{\unitlength}{1mm}
        \newcount\xco
        \newcount\yco
        \xco=-21
        \yco=12
        \begin{picture}(140,0)
        \put(\xco,\yco){$\ktl$}
        \advance\yco by-1
        {\loop
        \put(\xco,\yco){$\kcr$}
        \advance\yco by-2
        \ifnum\yco>-240
        \repeat
        \put(\xco,\yco){$\kbl$}}
        \xco=158
        \yco=12
        \put(\xco,\yco){$\ktr$}
        \advance\yco by-1
        {\loop
        \put(\xco,\yco){$\kcr$}
        \advance\yco by-2
        \ifnum\yco>-240
        \repeat
        \put(\xco,\yco){$\kbr$}}
        \put(-20,13){\tiny **University of Maryland * Center for String and
         Particle  Theory* Physics Department***University of Maryland *Center
        for String and Particle  Theory** }
        \put(-20,-241.5){\tiny **University of Maryland * Center for String and
         Particle  Theory* Physics Department***University of Maryland *Center
        for String and Particle  Theory** }
        \end{picture}
        \par\vskip-8mm}

\def\headpic{                                           % UM heading
        \indent
        \setlength{\unitlength}{.4mm}
        \thinlines
        \par
        \begin{picture}(29,16)
        \put(165,16){\line(1,0){4}}
        \put(170,16){\line(1,0){4}}
        \put(180,16){\line(1,0){4}}
        \put(175,0){\line(1,0){4}}
        \put(180,0){\line(1,0){4}}
        \put(185,0){\line(1,0){4}}
        \put(169,0){\line(0,1){16}}
        \put(170,0){\line(0,1){16}}
        \put(179,0){\line(0,1){16}}
        \put(180,0){\line(0,1){16}}
        \put(184,0){\line(0,1){16}}
        \put(185,0){\line(0,1){16}}
        \put(169,16){\oval(8,32)[bl]}
        \put(170,16){\oval(8,32)[br]}
        \put(179,0){\oval(8,32)[tl]}
        \put(185,0){\oval(8,32)[tr]}
        \end{picture}
        \par\vskip-6.5mm
        \thicklines}
\def\endtitle{\end{quotation}\newpage}                  % end title page

\newskip\humongous \humongous=0pt plus 1000pt minus 1000pt
\def\caja{\mathsurround=0pt}
\def\eqalign#1{\,\vcenter{\openup2\jot \caja
        \ialign{\strut \hfil$\displaystyle{##}$&$
        \displaystyle{{}##}$\hfil\crcr#1\crcr}}\,}
\newif\ifdtup

\begin{document}

\def\dt#1{\on{\hbox{\bf .}}{#1}}                % (big) dot over
\def\Dot#1{\dt{#1}}

\def\gfrac#1#2{\frac {\scriptstyle{#1}}
        {\mbox{\raisebox{-.6ex}{$\scriptstyle{#2}$}}}}
\def\gg{{\hbox{\sc g}}}
\border\headpic {\hbox to\hsize{\today \hfill
{UMDEPP-014-009}}}
\par \noindent
{ \hfill
%{hep-th/xxxx.xxxx}
}
\par

\setlength{\oddsidemargin}{0.3in}
\setlength{\evensidemargin}{-0.3in}
\begin{center}
{\large\bf Adinkra `Color' Confinement
\vskip.025in In Exemplary Off-Shell 
Constructions Of 
\vskip.03in
4D, ${\cal N}$ = 2 Supersymmetry Representations}\\[.3in]
S.\, James Gates, Jr.\footnote{gatess@wam.umd.edu}${}^{\dagger}$
and Kory Stiffler\footnote{kmstiffl@iun.edu}${}^*$
\\[0.15in]
${}^\dag${\it Center for String and Particle Theory\\
Department of Physics, University of Maryland\\
College Park, MD 20742-4111 USA}
\\[0.15in] 
and
\\[0.15in] 
${}^*${\it Department of Chemistry, Physics, and Astronomy\\
Indiana University Northwest\\
Gary, Indiana 46408 USA}
\\[.6in]
{\bf ABSTRACT}\\[.01in]
\end{center}
\begin{quotation}
{Evidence is presented  in some examples that an adinkra quantum number, 
$\chi_{\rm o}$ (arXiv:\ 0902.3830 [hep-th]), seems to play a role with regard 
to off-shell 4D, $\cal N$ = 2 SUSY similar to the role of color in QCD.  The vanishing of 
this adinkra quantum number appears to be a condition required for when 
two off-shell 4D, $\cal N$ = 1 supermultiplets form an off-shell 4D, $\cal N$ 
= 2 supermultiplet.  We also explicitly comment on a deformation of the Lie 
bracket and anti-commutator operators that has been extensively and implicitly
used in our 
work on ``Garden Algebras'' adinkras, and codes.}
\\[.5in]
\vfill
\endtitle

\setlength{\oddsidemargin}{0.3in}
\setlength{\evensidemargin}{-0.3in}

\setcounter{equation}{0}
\section{Introduction}

$~~~$ For a number of years, we have been developing a  
``Garden Algebra,'' Adinkras, and  Codes \cite{GRana} - 
\cite{Bowtie} approach for providing a deeper understanding 
of puzzling aspects of why and how supersymmetric {\em {off-shell}} 
representation theory works as it does.  The basis of this 
is our conjecture that {\em {off-shell}}  supersymmetrical 
representation theory should share as many as possible 
features with the representation theory of Lie compact algebras, 
but must be distinctive is some ways.  Let us use $su(3)$ as 
an exemplar of the former.

For $su(3)$ it is well accepted that the fundamental representations, 
provided by quark  triplets and anti-quark  triplets,  give the basic 
degrees of freedom from which to understand all representations 
of baryonic matter.  In the work of Ref.~\cite{G-1} , it was proposed 
that there exist fundamental objects, given the names `cis-adinkras' 
and `trans-adinkras,' which play a similar role in the context of 4D, 
$\cal N$ $=$ 1 supermultiplet representations.  However, within 
the work of \cite{permutadnk}, it was shown that there is a degeneracy 
in the `trans-adinkras,' that can be recognized by considering the
representation theory of the permutation group ${\cal S}_4$ and which is embedded in all adinkras with more
than four colors.  This allows the imposition of an intrinsic
class structure on adinkras and these classes become relevant
for defining how adinkras are related to the higher dimensional
supermultiplets. 

So the final result of our analysis is that there are three distinct 
off-shell adinkra classes, which can be identified with the 4D, 
$\cal N$ = 1 chiral, vector and tensor supermultiplets, that are 
the irreducible supersymmetry or SUSY equivalent to quarks.  
Just as all hadronic matter can be regarded as composites of 
$p$ quark triplets and $q$ anti-quark triplets (here $p$ and $q$ 
are simply integers), our research suggests all {\em {off}}-{\em 
{shell}} 4D, $\cal N$ $=$ 1 supermultiplets may be regarded 
as composites of $p$ chiral valise adinkras, $q$ vector valise 
adinkras and  $r$ tensor valise adinkras (where $p$, $q$, 
and $r$ are integers).

Although the Quark Model is now well accepted as being of fundamental
importance to describing hadronic matter and its interactions, it is often 
forgotten that one of the major reasons the Quark Model was initially 
accepted had to do with the discovery of the $\O{}^-$ composed of three 
strange quarks and first seen in 1964.  This particle had been predicted 
by the Quark Model prior to being seen in the laboratory and was a true 
`smoking gun' indicating the validity of the Quark Model as an accurate 
description of physics in Nature. Furthermore, analysis of the statistics of composites in
the Quark Model led to the discovery of color~\cite{Color}, the fiftieth 
anniversary of which has most recently been celebrated.

In this work, we provide more compelling evidence to support the 
assertion about an adinkra-based model of {\em {off-shell}} 
supersymmetric 4D, $\cal N$ = 1 representations.  Of course, 
since SUSY has not  yet been seen in the laboratory, our 
evidence must perforce be purely mathematical.  An assertion 
such as we have made ought to have implications for the structure 
of off-shell SUSY representations that go beyond the simple 
one-dimensional context used to discover the three foundational 
adinkras.  We demonstrate one such implication in this work.

 %%%%%%%%%%%%%%%%%%%%%%%%%%%%%%%%%%%%%%%%%%
\section{A Mathematical Background  Question}

$~~~~$ Some time ago, the topic of $q$-deformation of the usual Lie
bracket
\be
{ \big[  } \,  A ~,~ B \, {\big ]}   ~=~ A B ~-~ BA 
~\to ~
{ \big[  } \,  A ~,~ B \, {\big ]}_q   ~=~ A B ~-~ q \, BA  ~~,
\ee
where $-1  \le q \le 1$, was subject to numbers of studies (see for example
\cite{Q}).  We have not formally commented previously, but in some
ways the work of \cite{GRana} can be interpreted in a similar manner.
One can imagine two operators defined by
\be  \eqalign{
{ \big[  } \,  A ~,~ B \, {\big ]}_{q{\cal {GR}}>}   ~&=~ A\,  (B^{\rm T}) ~-~ q \, B \,
(A^{\rm T})  ~~~~~, \cr
{ \big[  } \,  A ~,~ B \, {\big ]}_{q{\cal {GR}}<}   ~&=~  (A^{\rm T})  \,  B ~-~ q \, 
 (B^{\rm T})  \, A   ~~~~~,
} \label{qGR} \ee
acting on matrices $A$ and $B$.  The quantities $ (A^{\rm T})$ and $(B^{\rm T})$
correspond to the respective transposed matrices.  This bracket (for
$q$ = $-1$) has shown up as part of the mathematical foundation of the
structures we call the `Garden Algebras.'  It is an interesting question
(to which we do not possess an answer) as whether our use of such
brackets can be extended in other ways?  One such possibility would
be to ask whether such a bracket admits analogs of Lie algebras?

To use matrices in such a construction we would begin with some set $\{ \cal G \}$ 
with $N$ elements denoted by $g_1$, $g_2$, $\dots$, $g_N$ and impose 
upon them the conditions
\be  \eqalign{
{ \big[  } \,  A ~,~ B \, {\big ]}_{q{\cal {GR}}>}   ~&=~ i \, f^{>}_{A \, B}{}^C \, h^{>}_C
 ~~~~~, \cr
{ \big[  } \,  A ~,~ B \, {\big ]}_{q{\cal {GR}}<}   ~&=~  i \, f^{<}_{A \, B}{}^C \, h^{<}_C
 ~~~~~,
} \ee
where $ h^{>}_C$ and $ h^{<}_C$ are other matrices and $ f^{>}_{A \, B}{}^C$
and $ f^{<}_{A \, B}{}^C$ are analogous to structure constants.  One other
feature of the ``q${\cal {GR}}$'' brackets in (\ref{qGR}) is that they permit non-diagonal
matrices to be used in their calculations.  Thus, if $A$ is a d${}_L$ $\times$ d${}_R$
and $B$ is a d${}_R$ $\times$ d${}_L$ matrix, then $h^{>}_C$ will be a d${}_L$ 
$\times$ d${}_L$ matrix and $h^{<}_C$ will be a d${}_R$ $\times$ d${}_R$ matrix.
We have long used these properties on our previous works investigating Garden
Algebras, adinkras, and codes.

 %%%%%%%%%%%%%%%%%%%%%%%%%%%%%%%%%%%%%%%%%%
\section{Building $\bm {\cal N}$ $=$ 2 Supermultiplets From 
$\bm {\cal N}$ $=$ 1 Supermultiplets}

$~~~~$ The basic idea of `Garden Algebras' is very simple.  The bracket operations
above are used to impose upon the N elements Clifford algebra-like conditions
\be
{ \big[  } \,  g_A ~,~ g_B \, {\big ]}_{(-1){\cal {GR}}>}     ~=~ 2 \, \d_{A \, B}\,  {\rm I}  
~~~~,~~~~
{ \big[  } \,  g_A ~,~ g_B \, {\big ]}_{(-1){\cal {GR}}<}   ~=~ 2 \, \d_{A \, B}\,  {\rm I}  
~~,~~
\ee
where $\rm I$ is the identity map and both equations are valid for all values of 
$A$ and $B$.  When the elements satisfy these conditions, we say $g_1$, $g_2$, 
$\dots$, $g_N$ forms a ``Garden Algebra.''  The set $\{ \cal G \}$ can be subject
to the more stringent requirement that it form a group.  However, this is not a
requirement. There is no a priori choices made for the group $\cal G$.  The 
question of whether all groups allow non-vanishing solutions to these conditions 
is unknown.  However, when  $\cal G$ is picked to be one of the orthogonal 
groups ${\rm O}$(d), these have been found to be important for the representations 
of space-time supersymmetry realized off-shell.

The Garden Algebras (GA's) go back the the oldest part of our system of 
analysis \cite{GRana} and when coupled with the SUSY Holography conjecture 
\cite{ENUF} assert {\em {all}} supermultiplets that are off-shell and possess no central 
charges must be representations to which the 0-brane reduction leads to a 
set of matrices $ (\,{\rm L}_{\rm I}\,)$ that satisfy
\be 
\eqalign{
 (\,{\rm L}_{\rm I}\,)_i{}^\hj\>(\,{\rm R}_{\rm J}\,)_\hj{}^k + (\,{\rm L}_{\rm J}\,)_i
 {}^\hj\>(\,{\rm R}_{\rm I}\,)_\hj{}^k &= 2\,\d_{{\rm I}{\rm J}}\,\d_i{}^k~~,\cr
(\,{\rm R}_{\rm J}\,)_\hi{}^j\>(\, {\rm L}_{\rm I}\,)_j{}^\hk + (\,{\rm R}_{\rm I}\,
)_\hi{}^j\>(\,{\rm L}_{\rm J}\,)_j{}^\hk &= 2\,\d_{{\rm I}{\rm J}}\,\d_\hi{}^\hk~~,
}  \label{GarDNAlg1}
\ee
\be
~~~~ (\,{\rm R}_{\rm I}\,)_\hj{}^k\,\d_{ik} = (\,{\rm L}_{\rm I}\,)_i{}^\hk\,\d_{\hj\hk}~~,
\label{GarDNAlg2}
\end{equation}
which we have denoted as the ``$\cal {GR}$(d, N) Algebras.''  Here the indices 
have ranges that correspond to $\rm I$, $\rm J$, $\dots$  = 1, $\dots$, N; i, j, $\dots$
= 1, $\dots$, d; and $\hi$, $\hj$, $\dots$ =  1, $\dots$, d for some integers N, and d.  
For this paper, we mostly consider the cases of N = 8 (for 4D, $\cal N$ $=$ 2) 
and N = 4 (for 4D, $\cal N$ $=$ 2) 1d, SUSY.

However, there are closely related algebraic structures that we denote as the
``${ { {\cal GR} ({\rm d}_L, \, {\rm d}_R, \,  {\rm N})}}$ Algebras''  that satisfy
\be 
\eqalign{
 (\,{\rm L}_{\rm I}\,)_i{}^\hj\>(\,{\rm R}_{\rm J}\,)_\hj{}^k + (\,{\rm L}_{\rm J}\,
 )_i{}^\hj\>(\,{\rm R}_{\rm I}\,)_\hj{}^k
&= 2\,\d_{{\rm I}{\rm J}}\,\d_i{}^k    ~+~ \D{}_{{\rm I}{\rm J}}{}_i{}^k  ~~,\cr
(\,{\rm R}_{\rm J}\,)_\hi{}^j\>(\, {\rm L}_{\rm I}\,)_j{}^\hk + (\,{\rm R}_{\rm I}\,
)_\hi{}^j\>(\,{\rm L}_{\rm J}\,)_j{}^\hk
&= 2\,\d_{{\rm I}{\rm J}}\,\d_\hi{}^\hk ~+~ {\Hat \D}{}_{{\rm I}{\rm J}}{}_\hi{}^\hk
~~,
}  \label{GarDNAlg3}
\ee
\be
~~~~ (\,{\rm R}_{\rm I}\,)_\hj{}^k\,\d_{ik} = (\,{\rm L}_{\rm I}\,)_i{}^\hk\,\d_{\hj\hk}~~,
\label{GarDNAlg4}
\end{equation}
Here the indices have ranges that correspond to $\rm I$, $\rm J$, $\dots$  = 1, $\dots$, 
N; i, j, $\dots$ = 1, $\dots$, d${}_L$; and $\hi$, $\hj$, $\dots$ =  1, $\dots$, d${}_R$ for some 
integers $N$, d${}_L$, and d${}_R$ and for some quantities $\D{}_{{\rm I}{\rm J}}{}_i{}^k $
and ${\Hat \D}{}_{{\rm I}{\rm J}}{}_\hi{}^\hk$.   Past experience \cite{G-1} has shown us that
when there are off-shell central charges present in the higher dimensional theory, 
these cast `shadows' in the 1D models in the form of the non-vanishing values of 
the quantities  $\D{}_{{\rm I}{\rm J}}{}_i{}^k $ and ${\Hat \D}{}_{{\rm I}{\rm J}}{}_\hi{}^\hk$. 

The strategy of this section is to start with some well-known 4D, $ {\cal N}$ $=$ 1 
supermultiplets to explore the possibility of constructing 4D, $ {\cal N}$ $=$ 2 
supermultiplets.  The reason this works conceptually is described below.

Let some 4D, $ {\cal N}$ $=$ 1 supermultiplet denoted by $ \{ {\cal A} \} $ possess
an action $S_1({\cal A} {\big |} {\cal A})$ quadratic in its fields and invariant under the 
action of an off-shell SUSY operator D${}_a$.  This means that
\be
{\rm D}_a \, \left[ \, S_1({\cal A} {\big |} {\cal A}) \, \right] ~=~ 0 ~~~, ~~~ 
\ee
up to total derivative terms, and it is off-shell if the condition
\be \left\{ 
\, {\rm D}_a ~ ,~{\rm D}_b \, \right\} ~=~ i \, 2 (\g^{\m}) {}_{a \, b}\, \pa_{\m}
\ee
is satisfied on all fields without regard to any field equations. A second such 4D, $ 
{\cal N}$ $=$ 1 supermultiplet, with the same number of degrees of freedom, 
denoted by $ \{ {\cal B} \} $ will possess its own invariant action $S_2({\cal B} {\big |} 
{\cal B})$ that satisfies the same property.  Denoting the SUSY operator above by D${}^1_a$, it follows that D${}^1_a$
satisfies
\be
{\rm D}_a^1 \, \left[ \, S_1({\cal A} {\big |} {\cal A}) ~+~ S_2({\cal B} {\big |} {\cal B})
\, \right] ~=~ 0 ~~~,
\ee
from its linearity.  However, this statement guarantees that a second invariance
generated by an operator D${}^2_a$ satisfying
\be
{\rm D}_a^2 \, \left[ \, S_1({\cal A} {\big |} {\cal A}) ~+~ S_2({\cal B} {\big |} {\cal B})
\, \right] ~=~ 0 ~~~,
\ee
must also exist. The realization of this second operator is such that it maps the
bosons in the $ \{ {\cal A} \} $ supermultiplet into the fermions of the $ \{ {\cal B} \}$ 
(using the same equations as were the case of the fermions in the $\cal A$-multiplet)
and  maps the fermions in the $ \{ {\cal A} \} $ supermultiplet into the bosons of 
the $ \{ {\cal B} \} $ (using the same equations as were the case of the fermions 
in the $\cal B$-multiplet) that allowed the realization
of  D${}^1_a$.

So we have a second fermionic invariance, but is it an on-shell or off-shell supersymmetry?
To answer this requires calculating the anticommutator algebra of D${}^1_a$.
and D${}^2_a$ on all the component fields.  There is nothing in the above construction
that guarantees that the two fermionic generators must form an extended
off-shell 4D, $ {\cal N}$ $=$ 2 supersymmetry algebra and one must check on a
case-by-case basis.  In the following, we will carry out such checks using the
familiar 4D, $ {\cal N}$ $=$ 1 chiral, vector, and tensor supermultiplets to play the
roles of $ \{ {\cal A} \} $ and $ \{ {\cal B} \} $.  We will show that an interesting
dichotomy emerges.

%%%%%%%%%%%%%%%%%%%%%%%%%%%%%%%%%%%%%%%%%%
 %%%%%%%%%%%%%%%%%%%%%%%%%%%%%%%%%%%%%%%%%%
 %%%%%%%%%%%%%%%%%%%%%%%%%%%%%%%%%%%%%%%%%%
\subsection{Building an $ {\cal N}$ $=$ 2 Supermultiplet From 
Chiral $+$ Chiral $ {\cal N}$ $=$ 1 Supermultiplets}\label{s:CC}

$~~~~$  Among the first discussions of a 4D, $ {\cal N}$ $=$ 2 supermultiplet
containing only propagating field of spin-1/2 or less is the work by Fayet \cite{Fy8}
in which there appears a citation to a work by Wess \cite{Wss} as providing the initial
discussion of the `W-F hypermultiplet.'  For the sake of completeness we review
these results.

The transformation laws for the W-F hypermultiplet containing the Chiral-Chiral 
multiplet combination are (our notational conventions can be found in the \cite{G-1})
\begin{align}
\begin{split}
{\rm D}_a^i A &= (\sigma^3)^{ij}\psi_a^j ~~~~~~~~~~\,~~,~~ {\rm D}_a^i B = i 
(\gamma^5)_a^{~b} \psi_b^i ~~~~~~~~~~~~,\\
{\rm D}_a^i F &= (\sigma^3)^{ij}(\gamma^{\mu})_a^{~b} \partial_{\mu}\psi_b^j ~~,~~
{\rm D}_a^i G = i (\gamma^5\gamma^\mu)_a^{~b}\partial_\mu \psi_b^i ~~~~~~,   \\
  {\rm D}_a^i \tilde A &= (\sigma^1)^{ij}\psi_a^j ~~~~~~~~~~\,~~,~~ 
  {\rm D}_a^i \tilde B = -(\sigma^2)^{ij} (\gamma^5)_a^{~b} \psi_b^j~~\,~,\\
  {\rm D}_a^i \tilde F &= (\sigma^1)^{ij}(\gamma^{\mu})_a^{~b} \partial_{\mu}\psi_b^j ~~,~~
  {\rm D}_a^i \tilde G = -(\sigma^2)^{ij} (\gamma^5\gamma^\mu)_a^{~b}\partial_\mu \psi_b^j,\\
  {\rm D}_a^i \psi_b^j &= i(\sigma^3)^{ij}\left((\gamma^{\mu})_{ab}\partial_{\mu}A-C_{ab}F\right) + 
  \delta^{ij}(-(\gamma^5\gamma^{\mu})_{ab}\partial_{\mu}B + (\gamma^5_{ab}) G) \\
  &~~~+~ i(\sigma^1)^{ij}\left((\gamma^{\mu})_{ab}\partial_{\mu}\tilde A-C_{ab}\tilde F\right) +
  i(\sigma^2)^{ij}(-(\gamma^5\gamma^{\mu})_{ab}\partial_{\mu}\tilde B + (\gamma^5_{ab})\tilde G) ~~~,
 \end{split}
\end{align}
\noindent where $i = 1,2$ labels the two supersymmetries, %$(\vec \sigma)$ denote the usual Pauli matrices,
and
\begin{align}
   (\s^0)^{ij}=\delta^{ij} 
\end{align}
The following Lagrangian is invariant with respect to these transformations:
\begin{align}
   \mathcal{L} &= -\frac{1}{2} \partial_{\mu}A \partial^\mu A -\frac{1}{2} \partial_{\mu}\tilde{A} \partial^\mu \tilde{A}-\frac{1}{2} \partial_{\mu}B \partial^\mu B-\frac{1}{2} \partial_{\mu}\tilde{B} \partial^\mu \tilde{B} + \nonumber\\
&~+\frac{1}{2} F^2 + \frac{1}{2} \tilde{F}^2 + \frac{1}{2} G^2 + \frac{1}{2} \tilde{G}^2 +\frac{1}{2}i(\gamma^\mu)^{cd}\psi^i_c \partial_\mu \psi^i_d 
\end{align}
which is easily seen to be the direct sum of the $\cal N$ $=$ 1 invariant Lagrangians for
the separate ($A, \, \psi^1_c , \, F $) chiral supermultiplet and the (${\Tilde A}, \, \psi^2_c , 
\, {\Tilde F} $) chiral supermultiplet.  Direct calculation yields the following algebra:
\begin{align}\label{e:CCAlgebra}
\begin{split}
\left\{{\rm D}^i_a,{\rm D}^j_b\right\}A=&\delta^{ij}2i\left(\gamma^\mu\right)_{ab}\partial_\mu A+i\left(\sigma^2\right)^{ij}2iC_{ab}\tilde F\\
\left\{{\rm D}^i_a,{\rm D}^j_b\right\}\tilde A=&\delta^{ij}2i\left(\gamma^\mu\right)_{ab}\partial_\mu\tilde A-i\left(\sigma^2\right)^{ij}2iC_{ab}F\\
\left\{{\rm D}^i_a,{\rm D}^j_b\right\}B=&\delta^{ij}2i\left(\gamma^\mu\right)_{ab}\partial_\mu B+i\left(\sigma^2\right)^{ij}2iC_{ab}\tilde G\\
\left\{{\rm D}^i_a,{\rm D}^j_b\right\}\tilde B=&\delta^{ij}2i\left(\gamma^\mu\right)_{ab}\partial_\mu\tilde B-i\left(\sigma^2\right)^{ij}2iC_{ab} G\\
\left\{{\rm D}^i_a,{\rm D}^j_b\right\}F=&\delta^{ij}2i\left(\gamma^\mu\right)_{ab}\partial_\mu F+i\left(\sigma^2\right)^{ij}2iC_{ab}\square\tilde A\\
\left\{{\rm D}^i_a,{\rm D}^j_b\right\}\tilde F=&\delta^{ij}2i\left(\gamma^\mu\right)_{ab}\partial_\mu\tilde F-i\left(\sigma^2\right)^{ij}2iC_{ab}\square A\\
\left\{{\rm D}^i_a,{\rm D}^j_b\right\}G=&\delta^{ij}2i\left(\gamma^\mu\right)_{ab}\partial_\mu G+i\left(\sigma^2\right)^{ij}2iC_{ab}\square\tilde B\\
\left\{{\rm D}^i_a,{\rm D}^j_b\right\}\tilde G=&\delta^{ij}2i\left(\gamma^\mu\right)_{ab}\partial_\mu\tilde G-i\left(\sigma^2\right)^{ij}2iC_{ab}\square B\\
\left\{{\rm D}^i_a,{\rm D}^j_b\right\}\psi^k_c=&\delta^{ij}2i\left(\gamma^\mu\right)_{ab}\partial_\mu \psi^k_c-\left(\sigma^2\right)^{ij}\left(\sigma^2\right)^{kr}2iC_{ab}\left(\gamma^\mu\right)_{c}^{~d}\partial_\mu\psi^r_d
\end{split}
%\label{eq:var}
\end{align}

From the results in Eqs.~(\ref{e:CCAlgebra}) we can see there must be an additional
symmetry of the Lagrangian with respect to the variations:
\begin{align}
   \delta A = P \tilde{F},&~~~\mbox{with}~~~ \delta \tilde{F} = -P \square A \\          
   \delta\tilde{A} = P F,&~~~\mbox{with}~~~ \delta F = - P \square\tilde{A} \\ 
   \delta B = P \tilde{G},&~~~\mbox{with}~~~ \delta\tilde{G} = -P \square B \\
   \delta\tilde{B} = P G,&~~~\mbox{with}~~~ \delta G = - P\square\tilde{B} \\
   \delta \psi^k_c &= P (\sigma^2)^{kr}(\gamma^\mu)_c^{~d} \,\partial_\mu \psi^r_d
\end{align}
\noindent where $P$ is a constant parameter.  In Eqs.~(\ref{e:CCAlgebra}) is
the composition of the two supersymmetry variation parameter according to
\begin{align}
   P &\equiv \varepsilon^a_i \varepsilon^b_j (\sigma^2)^{ij} C_{ab}
\end{align}
\noindent where $\varepsilon^a_i$ is an infinitesimal Grassmann spinor.  

When the fields ($A, \, {\Tilde A}, \, B, \, {\Tilde B}, \, F, \, {\Tilde F}, \,
G, \, {\Tilde G}, \,  \psi^r_a$) satisfy their equations of motion, all the variations
in (17) - (21) vanish.  As well, the `extra terms' in the anticommutators of (16) also
vanish.  So the symmetry generated by these variations are only non-trivial
off the mass shell.  This led to these being named as ``off-shell central charges.''
  
We now dimensionally reduce to an eight by eight adinkra by considering all 
fields to have only temporal dependence. As in~\cite{N4Off}, we identify
\begin{align}
   {\psi}^1_1 &= i {\Psi}_1,~~~{\psi}^1_2 = i {\Psi}_2,~~~{\psi}^1_3 = i {\Psi}_3,~~~{\psi}^1_4 = i{\Psi}_4,\nonumber\\
   \psi^2_1 &= i {\Psi}_5,~~~\psi^2_2 = i{\Psi}_6,~~~\psi^2_3 = i{\Psi}_7,~~~\psi^2_4 = i {\Psi}_8, \nonumber\\
   {\Phi}_1 &= A,~~~{\Phi}_2 = B,~~~\partial_0{\Phi}_3 = F,~~~\partial_0{\Phi}_4 = G,\nonumber\\
   {\Phi}_5 &= \tilde A,~~~{\Phi}_6 = \tilde B,~~~\partial_0{\Phi}_7 = \tilde F,~~~\partial_0{\Phi}_8 = \tilde G,
\end{align} 
\noindent and define
\begin{align}
\label{e:Ddef}
{\rm D}_{\rm I}=
 \begin{cases} 
 {\rm D}^1_{\rm I}&1\leq {\rm I}\leq4\\
 {\rm D}^2_{{\rm I}-4}&5\leq {\rm I}\leq8
 \end{cases}
\end{align}
\noindent whereupon the supersymmetric transformations reduce to
\begin{align}
   {\rm D}_{\rm I}{\Phi}_j &= i({\rm L}_{\rm I})_{j\hat{k}}{\Psi}_{\hat{k}}~~~,~~~{\rm D}_{\rm I} {\Psi}_{\hat{k}} = ({\rm R}_{\rm I})_{\hat{k}j}\partial_0 {\Phi}_j.
   \label{eq:CCLR}
\end{align}
The explicit form of the matrices in these equations are given in Appendix A.
These matrices satisfy the orthogonal relationship
\begin{align}\label{e:orthogonal}
   \bm{\rm L}_{\rm I} = (\bm{\rm R}_{\rm I})^{-1} = (\bm{\rm R}_{\rm I})^{\text{T}}
\end{align}
The Chromocharacters are defined as 
\begin{align}
\label{}
\begin{split}
\left(\varphi^{(p)}\right)_{{\rm I}_1{\rm J}_1\dots {\rm I}_p {\rm J}_p}=&\text{Tr}\left\{{\rm L}_{{\rm I}_1}\left({\rm L}_{{\rm J}_1}\right)^{\text{T}}\cdots {\rm L}_{{\rm I}_p}\left({\rm L}_{{\rm J}_p}\right)^{\text{T}}\right\}\\
\left(\tilde\varphi^{(p)}\right)_{{\rm I}_1{\rm J}_1\dots {\rm I}_p{\rm J}_p}=&\text{Tr}\left\{\left({\rm L}_{{\rm I}_1}\right)^{\text{T}}{\rm L}_{{\rm J}_1}\cdots \left({\rm L}_{{\rm I}_p}\right)^{\text{T}}{\rm L}_{{\rm J}_p}\right\}
\end{split}
\label{Cchrm}
\end{align}
\noindent with first order chromocharacters given by:
\begin{align}
(\varphi^{(1)})_{{\rm I}_1 {\rm J}_1} &= (\tilde{\varphi}^{(1)})_{{\rm I}_1 {\rm J}_1} = 8\delta_{{\rm I}_1 {\rm J}_1}
\end{align}
\noindent and second order characters given by:
\begin{align}\label{e:CC2Chromos}
\begin{split}
(\varphi^{(2)})_{{\rm I}_1 {\rm J}_1 {\rm I}_2 {\rm J}_2} ~=~&8 \left(\delta _{{\rm I}_1 {\rm J}_1} \delta _{{\rm I}_2 {\rm J}_2}+\left(\sigma^0 \otimes \sigma^3
   \otimes \sigma ^2\right){}_{{\rm I}_1 {\rm J}_1} \left(\sigma^0 \otimes \sigma^3
   \otimes \sigma ^2\right){}_{{\rm I}_2 {\rm J}_2}\right.\\
   &+~\left.\left(\sigma^3 \otimes \sigma
   ^2\otimes \sigma^0\right){}_{{\rm I}_1 {\rm J}_1} \left(\sigma^3 \otimes \sigma ^2\otimes
   \sigma^0\right){}_{{\rm I}_2 {\rm J}_2}\right.\\
   &+~\left.\left(\sigma ^3\otimes \sigma ^1\otimes \sigma ^2\right){}_{{\rm I}_1
   {\rm J}_1} \left(\sigma ^3\otimes \sigma ^1\otimes \sigma ^2\right){}_{{\rm I}_2
   {\rm J}_2}\right.\\
   &+~\left.\left(\sigma ^2\otimes \sigma ^0\otimes \sigma ^0\right){}_{{\rm I}_1 {\rm J}_1}
   \left(\sigma ^2\otimes \sigma ^0\otimes \sigma ^0\right){}_{{\rm I}_2
   {\rm J}_2}\right.\\
   &+~\left.\left(\sigma ^2\otimes \sigma ^3\otimes \sigma ^2\right){}_{{\rm I}_1
   {\rm J}_1} \left(\sigma ^2\otimes \sigma ^3\otimes \sigma ^2\right){}_{{\rm I}_2
   {\rm J}_2}\right.\\
   &+~\left.\left(\sigma ^1\otimes \sigma^2 \otimes \sigma ^0\right){}_{{\rm I}_1
   {\rm J}_1} \left(\sigma ^1\otimes \sigma^2 \otimes \sigma ^0\right){}_{{\rm I}_2
   {\rm J}_2}\right.\\
   &+~\left.\left(\sigma ^1\otimes \sigma^1 \otimes \sigma ^2\right){}_{{\rm I}_1 {\rm J}_1}
   \left(\sigma ^1\otimes \sigma^1 \otimes \sigma ^2\right){}_{{\rm I}_2 {\rm J}_2}\right)\\
(\tilde\varphi^{(2)})_{{\rm I}_1 {\rm J}_1 {\rm I}_2 {\rm J}_2} ~=~&8 \left(\delta _{{\rm I}_1 
{\rm J}_1} \delta _{{\rm I}_2 {\rm J}_2}+\left(\sigma ^2\otimes \sigma^3 \otimes \sigma^2\right){
}_{{\rm I}_1 {\rm J}_1} \left(\sigma ^2\otimes \sigma^3 \otimes \sigma ^2\right){}_{{\rm I}_2 {\rm 
J}_2}\right.\\
   &+~\left.\left(\sigma ^0\otimes \sigma^0 \otimes \sigma ^2
   \right){}_{{\rm I}_1 {\rm J}_1} \left(\sigma ^0\otimes \sigma^0 \otimes \sigma ^2
   \right){}_{{\rm I}_2 {\rm J}_2}\right.\\
   &+~\left.\left(\sigma ^0\otimes \sigma^2 \otimes \sigma^3
   \right){}_{{\rm I}_1 {\rm J}_1} \left(\sigma ^0\otimes \sigma^2 \otimes \sigma^3
   \right){}_{{\rm I}_2 {\rm J}_2}\right.\\
   &+~\left.\left(\sigma ^0\otimes \sigma^2 \otimes \sigma^1
   \right){}_{{\rm I}_1 {\rm J}_1} \left(\sigma ^0\otimes \sigma^2 \otimes \sigma^1\right){}_{{\rm 
   I}_2 {\rm J}_2}\right.\\
   &+~\left.\left(\sigma ^2\otimes \sigma^3 \otimes \sigma
   ^0\right){}_{{\rm I}_1 {\rm J}_1} \left(\sigma ^2\otimes \sigma^3 \otimes \sigma
   ^0\right){}_{{\rm I}_2 {\rm J}_2}\right.\\
   &+~\left.\left(\sigma ^2\otimes \sigma^1 \otimes \sigma^3\right){}_{{\rm I}_1 {\rm 
   J}_1} \left(\sigma ^2\otimes \sigma^1 \otimes \sigma^3\right){}_{{\rm I}_2 {\rm J}_2}\right.\\
   &+~\left.\left(\sigma ^2\otimes \sigma ^1\otimes \sigma
   ^1\right){}_{{\rm I}_1 {\rm J}_1} \left(\sigma ^2\otimes \sigma ^1\otimes \sigma
   ^1\right){}_{{\rm I}_2 {\rm J}_2}\right)
\end{split}
\end{align}

Finally the L-matrices and R-matrices that arise in the case of combining two
4D $\cal N$ $=$ 1 chiral supermultiplets in an attempt to derive a 4D $\cal N$ $=$ 2 supermultiplet
satisfy (\ref{GarDNAlg3}) where,
\begin{align}
\D{}_{{\rm I}{\rm J}}{}_i{}^k  
~=~ - \, 2\, \left(\sigma^2 \otimes \sigma^3\otimes \sigma^2\right){}_{{\rm I}{\rm J}} 
\left(\sigma^2\otimes \sigma^2\otimes \sigma^0\right)_i^{~k}  ~~~,
\end{align}
\noindent and  
\begin{align}
{\Hat \D}{}_{{\rm I}{\rm J}}{}_\hi{}^\hk
~&=~ -\, 2 \, \left(\sigma^2 \otimes \sigma^3\otimes \sigma^2\right){}_{{\rm I}{\rm J}} 
\left(\sigma^2\otimes \sigma^3\otimes \sigma^2\right)_{\hat{i}}^{~\hat{k}} ~~~.
\end{align}

%%%%%%%%%%%%%%%%%%%%%%%%%%%%%%%%%%%%%%%%%%
 %%%%%%%%%%%%%%%%%%%%%%%%%%%%%%%%%%%%%%%%%%
 %%%%%%%%%%%%%%%%%%%%%%%%%%%%%%%%%%%%%%%%%%
\subsection{Building a $\bm {\cal N}$ $=$ 2 Supermultiplet From 
Chiral $+$ Vector $\bm {\cal N}$ $=$ 1 Supermultiplets}\label{s:CV}

$~~~~$  The same work by Fayet \cite{Fy8} also introduced the now familiar 4D,
$ {\cal N}$ $=$ 2 `vector multiplet.'  From these we derive the following realization
of for the D-algebra,
\be \eqalign{
{\rm D}_a^1 A ~&=~ {\psi}_a  ~~~, \cr
%%%%%%%%%%%%%%%%%%%%%%%%%%%%
{\rm D}_a^1 B ~&=~ i \, (\gamma^5){}_a{}^b \, {\psi}_b  ~~~, \cr
%%%%%%%%%%%%%%%%%%%%%%%%%%%%
{\rm D}_a^1 {\psi}_b ~&=~ i\, (\gamma^\mu){}_{a \,b}\,  \partial_\mu A 
~-~  (\gamma^5\gamma^\mu){}_{a \,b} \, \partial_\mu B ~-~ i \, C_{a\, b} 
\,F  ~+~  (\gamma^5){}_{ a \, b} G  ~~, \cr
%%%%%%%%%%%%%%%%%%%%%%%%%%%%
{\rm D}_a^1 F ~&=~  (\gamma^\mu){}_a{}^b \, \partial_\mu \, {\psi}_b   ~~~, \cr
%%%%%%%%%%%%%%%%%%%%%%%%%%%%
{\rm D}_a^1 G ~&=~ i \,(\gamma^5\gamma^\mu){}_a{}^b \, \partial_\mu \,  
{\psi}_b  ~~~.
} \label{chi1}
\ee

\be \eqalign{
{\rm D}_a^1 \, A{}_{\mu} ~&=~  (\gamma_\mu){}_a {}^b \,  \l_b  ~~~, {~~~ 
~~~~~~~~~~~~~~~~~~~~~~~~~~~~~~~~} \cr
%%%%%%%%%%%%%%%%%%%%%%%%%%%%
{\rm D}_a^1 \l_b ~&=~   - \,i \, \fracm 14 ( [\, \gamma^{\mu}\, , \,  \gamma^{\nu} 
\,]){}_a{}_b \, (\,  \partial_\mu  \, A{}_{\nu}    ~-~  \partial_\nu \, A{}_{\mu}  \, )
~+~  (\gamma^5){}_{a \,b} \,    {\rm d} ~~, {~~~~~~} \cr
%%%%%%%%%%%%%%%%%%%%%%%%%%%%
{\rm D}_a^1 \, {\rm d} ~&=~  i \, (\gamma^5\gamma^\mu){}_a {}^b \, 
\,  \partial_\mu \l_b  ~~~. \cr
} \label{V1}
\ee

\be \eqalign{
{\rm D}_a^{2} A ~&=~ \l_a  ~~~, \cr
%%%%%%%%%%%%%%%%%%%%%%%%%%%%
{\rm D}_a^{2} B ~&=~ i \, (\gamma^5){}_a{}^b \, \l_b  ~~~, \cr
%%%%%%%%%%%%%%%%%%%%%%%%%%%%
{\rm D}_a^{2} \l_b ~&=~ i\, (\gamma^\mu){}_{a \,b}\,  \partial_\mu A 
~-~ (\gamma^5\gamma^\mu){}_{a \,b} \, \partial_\mu B ~-~ i \, C_{a\, b} 
\,F  ~-~ (\gamma^5){}_{ a \, b} G  ~~, \cr
%%%%%%%%%%%%%%%%%%%%%%%%%%%%
{\rm D}_a^{2} F ~&=~ (\gamma^\mu){}_a{}^b \, \partial_\mu \, \l_b   ~~~, \cr
%%%%%%%%%%%%%%%%%%%%%%%%%%%%
{\rm D}_a^{2} G ~&=~- i \,(\gamma^5\gamma^\mu){}_a{}^b \, \partial_\mu \,  
\l_b  ~~~.
} \label{chi1b}
\ee

\be \eqalign{
{\rm D}_a^{2} \, A{}_{\mu} ~&=~ - (\gamma_\mu){}_a {}^b \,  {\psi}_b  ~~~, \cr
%%%%%%%%%%%%%%%%%%%%%%%%%%%%
{\rm D}_a^{2} {\psi}_b ~&=~  \,i \, \fracm 14 ( [\, \gamma^{\mu}\, , \,  \gamma^{\nu} 
\,]){}_a{}_b \, (\,  \partial_\mu  \, A{}_{\nu}    ~-~  \partial_\nu \, A{}_{\mu}  \, )
~+~ (\gamma^5){}_{a \,b} \,    {\rm d} ~~, {~~~~~~~~~~}  \cr
%%%%%%%%%%%%%%%%%%%%%%%%%%%%
{\rm D}_a^{2} \, {\rm d} ~&=~ i \, (\gamma^5\gamma^\mu){}_a {}^b \, 
\,  \partial_\mu {\psi}_b  ~~~. \cr
} \label{V1b}
\ee
that are equivalent to an invariance, up to total derivatives, of the Lagrangian
\be\eqalign{
 \mathcal{L} = &  -\frac{1}{2} \partial_{\mu}A \partial^\mu A -\frac{1}{2} \partial_{\mu}B \partial^\mu B- \frac{1}{4} F_{\mu\nu} F^{\mu\nu} + \frac{1}{2} {\rm d}^2    \cr
 &+ \frac{1}{2} i (\gamma^\mu)^{bc} \l_b \partial_\mu \l_c +\frac{1}{2} i (\gamma^\mu)^{bc} {\psi}_b \partial_\mu {\psi}_c 
}\ee
where
\begin{align}
 F_{\mu\nu} = \partial_{\mu}A_{\nu} - \partial_{\nu}A_{\mu}~~~.
\end{align}
The transformation laws satisfy the algebra
\begin{align}
   \{ {\rm D}_a^i, {\rm D}_b^j\} \chi &= 2 i \d^{ij} (\gamma^{\mu})_{ab} \chi, \nonumber\\*
   \{{\rm D}_a^i, {\rm D}_b^j\}A_{\nu} &= 2 i \d^{ij}(\gamma^{\mu})_{ab} F_{\mu\nu}+ i(\sigma^2)^{ij}(2 i C_{ab} \partial_{\nu}A - 2(\gamma^5)_{ab} \partial_{\nu}B)
\end{align}
where
\be
  \chi = \left\{ A,B,F,G,d,{\psi}_c,\lambda_c \right\}~~~.
\ee
Note, the algebra closes up to gauge transformations.
We now dimensionally reduce to an eight by eight adinkra by considering all fields to have only temporal dependence. As in \cite{G-1}, we choose the gauge
\be
  A_0 = 0
\ee
and identify
\begin{align}
   {\psi}_1 &= i {\Psi}_1,~~~{\psi}_2 = i {\Psi}_2,~~~{\psi}_3 = i {\Psi}_3,~~~{\psi}_4 = i{\Psi}_4,\nonumber\\
   \lambda_1 &= i {\Psi}_5,~~~\lambda_2 = i{\Psi}_6,~~~\lambda_3 = i{\Psi}_7,~~~\lambda_4 = i {\Psi}_8, \nonumber\\
   {\Phi}_1 &= A,~~~{\Phi}_2 = B,~~~\partial_0{\Phi}_3 = F,~~~\partial_0{\Phi}_4 = G,\nonumber\\
   {\Phi}_5 &= A_1,~~~{\Phi}_6 = A_2,~~~{\Phi}_7 = A_3,~~~\partial_0{\Phi}_8 = {\rm d},
\end{align}  
and define ${\rm D}_{\rm I}$ as in Eq.~(\ref{e:Ddef}) whereupon the supersymmetric transformations reduce to the familiar form, Eq.~(\ref{eq:CCLR}), where now the adinkra matrices are given in appendix B
and the $\left({\rm R}_{\rm I}\right)_{\hat{k}i}$ are given by the orthogonal relationship, Eq.~(\ref{e:orthogonal}), as in Sec.~\ref{s:CC}.
The adinkra matrices satisfy the Garden Algebra in (\ref{GarDNAlg1}).  The first and second order chromocharacters are given by
\begin{align}\label{e:ChromoOneCV}
(\varphi^{(1)})_{{\rm I}_1 {\rm J}_1} &= (\tilde{\varphi}^{(1)})_{{\rm I}_1 {\rm J}_1} = 8 \delta_{{\rm I}_1 {\rm J}_1}
\end{align}
\noindent and
\begin{align}\label{e:ChromoTwoCV}
 &(\varphi^{(2)})_{{\rm I}_1 {\rm J}_1 {\rm I}_2 {\rm J}_2} = (\tilde{\varphi}^{(2)})_{{\rm I}_1 {\rm J}_1 {\rm I}_2 {\rm J}_2} %\nonumber\\
 %&~~~~~~~~~~
 ~=~ 8 (\delta_{{\rm I}_1 {\rm J}_1}\delta_{{\rm I}_2 {\rm J}_2} + \delta_{{\rm I}_1 {\rm J}_2}\delta_{{\rm I}_2 {\rm J}_1}-\delta_{{\rm I}_1 {\rm I}_2}\delta_{{\rm J}_1 {\rm J}_2}).
\end{align}

%%%%%%%%%%%%%%%%%%%%%%%%%%%%%%%%%%%%%%%%%%
 %%%%%%%%%%%%%%%%%%%%%%%%%%%%%%%%%%%%%%%%%%
 %%%%%%%%%%%%%%%%%%%%%%%%%%%%%%%%%%%%%%%%%%
\subsection{Building an $\bm {\cal N}$ $=$ 2 Supermultiplet From 
Chiral $+$ Tensor $\bm {\cal N}$ $=$ 1 Supermultiplets}\label{s:CT}

$~~~~$ The ``4D, $\cal N$ $=$ 2 tensor'' multiplet (also known as the O(2) multiplet)
was first introduced by Wess \cite{Wss} and in our notation has a set of transformation laws 
of the form
\be \eqalign{
{\rm D}_a^1 A ~&=~ {\psi}_a  ~~~, \cr
%%%%%%%%%%%%%%%%%%%%%%%%%%%%
{\rm D}_a^1 B ~&=~ i \, (\gamma^5){}_a{}^b \, {\psi}_b  ~~~, \cr
%%%%%%%%%%%%%%%%%%%%%%%%%%%%
{\rm D}_a^1 {\psi}_b ~&=~ i\, (\gamma^\mu){}_{a \,b}\,  \partial_\mu A 
~-~  (\gamma^5\gamma^\mu){}_{a \,b} \, \partial_\mu B ~-~ i \, C_{a\, b} 
\,F  ~+~  (\gamma^5){}_{ a \, b} G  ~~, \cr
%%%%%%%%%%%%%%%%%%%%%%%%%%%%
{\rm D}_a^1 F ~&=~  (\gamma^\mu){}_a{}^b \, \partial_\mu \, {\psi}_b   ~~~, \cr
%%%%%%%%%%%%%%%%%%%%%%%%%%%%
{\rm D}_a^1 G ~&=~ i \,(\gamma^5\gamma^\mu){}_a{}^b \, \partial_\mu \,  
{\psi}_b  ~~~.
} \label{chi1c}
\ee
\be \eqalign{
{\rm D}_a^1 \varphi ~&=~ \chi_a  ~~~, \cr
%%%%%%%%%%%%%%%%%%%%%%%%%%%%
{\rm D}_a^1 B{}_{\mu \, \nu } ~&=~ -\, \fracm 14 ( [\, \gamma_{\mu}
\, , \,  \gamma_{\nu} \,]){}_a{}^b \, \chi_b  ~~~, \cr
%%%%%%%%%%%%%%%%%%%%%%%%%%%%
{\rm D}_a^1 \chi_b ~&=~ i\, (\gamma^\mu){}_{a \,b}\,  \partial_\mu \varphi 
~-~  (\gamma^5\gamma^\mu){}_{a \,b} \, \e{}_{\mu}{}^{\r \, \s \, \t}
\partial_\r B {}_{\s \, \t}~~.  {~~~~~~~~~~~~~~~~~~~~\,~~}
} \label{ten1}
\ee
\be \eqalign{
{~~~~~~~~~\,~~\,~~} &{\rm D}_a^2 A ~=~ -\chi_a  ~~~, \cr
%%%%%%%%%%%%%%%%%%%%%%%%%%%%
&{\rm D}_a^2 B ~=~ i \, (\gamma^5){}_a{}^b \, \chi_b  ~~~, \cr
%%%%%%%%%%%%%%%%%%%%%%%%%%%%
&{\rm D}_a^2 \chi_b ~=~ -i\, (\gamma^\mu){}_{a \,b}\,  \partial_\mu A 
~-~  (\gamma^5\gamma^\mu){}_{a \,b} \, \partial_\mu B ~-~ i \, C_{a\, b} 
\,F  ~+~  (\gamma^5){}_{ a \, b} G  ~, \cr
%%%%%%%%%%%%%%%%%%%%%%%%%%%%
&{\rm D}_a^2 F ~=~  (\gamma^\mu){}_a{}^b \, \partial_\mu \, \chi_b   ~~~, \cr
%%%%%%%%%%%%%%%%%%%%%%%%%%%%
&{\rm D}_a^2 G ~=~ i \,(\gamma^5\gamma^\mu){}_a{}^b \, \partial_\mu \,  
\chi_b  ~~~.
} \label{chi1d}
\ee
\be \eqalign{
{\rm D}_a^2 \varphi ~&=~ {\psi}_a  ~~~, \cr
%%%%%%%%%%%%%%%%%%%%%%%%%%%%
{\rm D}_a^2 B{}_{\mu \, \nu } ~&=~ \, \fracm 14 ( [\, \gamma_{\mu}
\, , \,  \gamma_{\nu} \,]){}_a{}^b \, {\psi}_b  ~~~, \cr
%%%%%%%%%%%%%%%%%%%%%%%%%%%%
{\rm D}_a^2 {\psi}_b ~&=~ i\, (\gamma^\mu){}_{a \,b}\,  \partial_\mu \varphi 
~+~  (\gamma^5\gamma^\mu){}_{a \,b} \, \e{}_{\mu}{}^{\r \, \s \, \t}
\partial_\r B {}_{\s \, \t}~~. {~~~~~~~~~~~~~~~~~~~~~~}
} \label{ten1b}
\ee
derived from the supersymmetry invariance, up to total derivatives, of the Lagrangian
\be\eqalign{
 \mathcal{L} = &  -\frac{1}{2} \partial_{\mu}A \partial^\mu A -\frac{1}{2} \partial_{\mu}B \partial^\mu B  - \frc{1}{3}H_{\mu\nu\alpha}H^{\mu\nu\alpha}- \frc{1}2 \partial_\mu \varphi \partial^\mu \varphi   \cr
 &+ \frac{1}{2} i (\gamma^\mu)^{bc} \chi_b \partial_\mu \chi_c +\frac{1}{2} i (\gamma^\mu)^{bc} {\psi}_b \partial_\mu {\psi}_c + \frac{1}{2} F^2 + \frac{1}{2} G^2
}\ee
where
\be
  H_{\mu\nu\alpha} = \partial_\m B_{\n\a} + \partial_\n B_{\a\m} + \partial_\a B_{\m\n}  ~~~.
\ee

We find the algebra closes up to gauge transformations on $B_{\mu\nu}$:
\begin{align}
   \{{\rm D}_a^i, {\rm D}_b^j\} X &= 2 i \delta^{ij} (\gamma^{\mu})_{ab} X, \nonumber\\
   \{{\rm D}_a^i, {\rm D}_b^j\}B_{\mu\nu} &= 2 i \delta^{ij}(\gamma^{\alpha})_{ab} H_{\alpha\mu\nu}+ \nonumber\\
 &~~~+ i(\gamma_{\mu}\partial_{\nu} -\gamma_{\nu}\partial_{\mu})_{ac}[(\sigma^{1})^{ij}\delta_{b}^{~c}A + (\sigma^{2})^{ij}(\gamma^5)_{b}^{~c}B - (\sigma^3)^{ij}\delta_{b}^{~c}\varphi]
\end{align}
\noindent where
\begin{align}
   X = \left\{A, B , F , G ,\varphi ,{\psi}_c ,\chi_c \right\}
\end{align}

We now dimensionally reduce to an eight by eight adinkra by considering all fields to have only temporal dependence. As in \cite{G-1}, we choose the gauge
\be
  B_{0i} = 0
\ee
and identify
\begin{align}
   {\psi}_1 &= i {\Psi}_1,~~~{\psi}_2 = i {\Psi}_2,~~~{\psi}_3 = i {\Psi}_3,~~~{\psi}_4 = i{\Psi}_4,\nonumber\\*
   \chi_1 &= i {\Psi}_5,~~~\chi_2 = i{\Psi}_6,~~~\chi_3 = i{\Psi}_7,~~~\chi_4 = i {\Psi}_8, \nonumber\\*
   {\Phi}_1 &= A,~~~{\Phi}_2 = B,~~~\partial_0{\Phi}_3 = F,~~~\partial_0{\Phi}_4 = G,\nonumber\\*
   {\Phi}_5 &= \varphi,~~~{\Phi}_6 = 2 B_{12},~~~{\Phi}_7 = 2 B_{23},~~~{\Phi}_8 = 2 B_{31},
\end{align} 
using again the definition~(\ref{e:Ddef}) for ${\rm D}_{\rm I}$, the transformation rules can be cast into the form, Eq.~(\ref{eq:CCLR}), with the adinkra matrices as in Appendix C
with $\left({\rm R}_{\rm I}\right)_{\hat{k}i}$ once again given by the orthogonality relationship, Eq.~(\ref{e:orthogonal}).
These matrices satisfy the Garden Algebra, (\ref{GarDNAlg1}), and have the first and second order chromocharacters, Eq.~(\ref{e:ChromoOneCV}) and Eq.~(\ref{e:ChromoTwoCV}), the same as for the chiral-vector multiplet of Section~\ref{s:CV}.

 %%%%%%%%%%%%%%%%%%%%%%%%%%%%%%%%%%%%%%%%%%
 %%%%%%%%%%%%%%%%%%%%%%%%%%%%%%%%%%%%%%%%%%
 %%%%%%%%%%%%%%%%%%%%%%%%%%%%%%%%%%%%%%%%%%
\subsection{Building an $\bm {\cal N}$ $=$ 2 Supermultiplet From 
Vector $+$ Vector $\bm {\cal N}$ $=$ 1 Supermultiplets}\label{s:VV}

$~~~~$ The three constructs discussed previously in this chapter are well known.
Their starting point may be regarded as utilizing the three 4D, ${\cal N}$ $=$ 1 chiral,
vector, and tensor supermultiplets as building blocks for models with a higher
degree of extended SUSY.  However, a little thought reveals there are more
similar constructions to explore.  Since there are three distinct 4D, ${\cal N}$ $=$ 1 
building blocks, there should be 6 ways in which one can attempt to realize
4D, ${\cal N}$ $=$ 2 multiplets.  We direct our attention to these other possibilities 
in this and the next subsection.

For the case of two 4D, ${\cal N}$ $=$ 1 vector multiplets, we introduce the 
transformation laws for this system as
$$ \eqalign{
{\rm D}_a^i \, A{}_{\mu} ~&=~  (\gamma_\mu){}_a {}^b \, b^{ij} \l_b^j  ~~~, 
{~~~~~~~~~~~~~~~~~~~~~~~~~~~} {~~~~~~~~~~~~~~~~~~~~~~~~~~~~}
%%%%%%%%%%%%%%%%%%%%%%%%%%%% 
} 
$$
\be \eqalign{
{\rm D}_a^i \, \tilde{A}{}_{\mu} ~&=~  (\gamma_\mu){}_a {}^b \, a^{ij} \l_b^j  ~~~, \cr
%%%%%%%%%%%%%%%%%%%%%%%%%%%%
{\rm D}_a^i \l_b^j ~&=~   b^{ij} \left(- \,i \, \fracm 14 ( [\, \gamma^{\mu}\, , \,  \gamma^{\nu} 
\,]){}_a{}_b \, (\,  \partial_\mu  \, A{}_{\nu}    ~-~  \partial_\nu \, A{}_{\mu}  \, )
~+~  (\gamma^5){}_{a \,b} \,    {\rm d} \right) +  \cr
&~~~+a^{ij} \left(- \,i \, \fracm 14 ( [\, \gamma^{\mu}\, , \,  \gamma^{\nu} 
\,]){}_a{}_b \, (\,  \partial_\mu  \, \tilde{A}{}_{\nu}    ~-~  \partial_\nu \, \tilde{A}{}_{\mu}  \, )
~+~  (\gamma^5){}_{a \,b} \,    \tilde{\rm d} \right) +  \cr
%%%%%%%%%%%%%%%%%%%%%%%%%%%%
{\rm D}_a^i \, {\rm d} ~&=~  i \, (\gamma^5\gamma^\mu){}_a {}^b \, b^{ij}
\,  \partial_\mu \l_b^j  ~~~, \cr
%%%%%%%%%%%%%%%%%%%%%%%%%%%%
{\rm D}_a^i \, \tilde{\rm d} ~&=~  i \, (\gamma^5\gamma^\mu){}_a {}^b a^{ij} \, 
\,  \partial_\mu \l_b^j  ~~~, \cr
} \label{VV1}
\ee
where $i,j$ = 1,2 and
\be\eqalign{
a^{ij} = & \cos a_0~(\s^1)^{ij} + i \sin a_0~(\s^2)^{ij}~~~, \cr
b^{ij} = & \cos b_0~ \delta^{ij} + \sin b_0 ~ (\s^3)^{ij}~~~.
}\ee
The transformation laws~(\ref{VV1}) lead to an invariance of the Lagrangian
\be\eqalign{
 \mathcal{L} = & - \frac{1}{4} F_{\mu\nu} F^{\mu\nu} - \frac{1}{4} \tilde{F}_{\mu\nu} \tilde{F}^{\mu\nu} + \frac{1}{2} i (\gamma^\mu)^{bc} \l_b^j \partial_\mu \l_c^j + \frac{1}{2} {\rm d}^2 + \frac{1}{2} \tilde{\rm d}^2 
}\ee
where
\be
  F_{\mu\nu} = \partial_\mu A_\nu - \partial_\nu A_\mu~~~,~~~\tilde{F}_{\mu\nu} = \partial_\mu \tilde{A}_\nu - \partial_\nu \tilde{A}_\mu
\ee  
and satisfy the algebra
$$  \eqalign{
\left\{ {\rm D}_a^i, {\rm D}_b^j \right\} A_\nu = &  2 i (\d^{ij} + (\s^3)^{ij} \sin 2 
b_0) (\g^\m)_{ab} F_{\m\n}    \cr
&+ 2 i \left(\s^1 \right)^{ij} \cos(a_0 + b_0) (\g^\m)_{ab} \tilde{F}_{\m\n}   \cr
&- i (\s^2)^{ij} \sin(a_0 - b_0) \e_\n^{~\m\a\b} (\g^5 \g_\m)_{ab} \tilde{F}_{\a\b}   \cr
&- 2 i \sin(a_0 - b_0) (\s^2)^{ij} \sin(a_0 - b_0) (\g^5\g_\n)_{ab} \tilde{\rm d}
~~~, \cr
%%%%%%%%%%%%%%%%%%%%%%%%%%%%%%%%%
\left\{ {\rm D}_a^i, {\rm D}_b^j \right\} \tilde{A}_\nu = &  2 i(\d^{ij} + (\s^3)^{ij} 
\sin 2 b_0) (\g^\m)_{ab} \tilde{F}_{\m\n}    \cr
&+ 2 i \left(\s^1 \right)^{ij} \cos(a_0 + b_0) (\g^\m)_{ab} F_{\m\n}   \cr
&+ i (\s^2)^{ij} \sin(a_0 - b_0) \e_\n^{~\m\a\b} (\g^5 \g_\m)_{ab} F_{\a\b}   \cr
&+ 2 i \sin(a_0 - b_0) (\s^2)^{ij} \sin(a_0 - b_0) (\g^5\g_\n)_{ab} {\rm d}~~~, 
{~~~~~~~~~~~~~~~~~~~~~~~~~~~~~~}  \cr
\left\{ {\rm D}_a^i, {\rm D}_b^j \right\}  \l_c^k = & \left(2 i \d^{ij} \d^{km} + i (\cos 2 a_0 + \cos 2 b_0 )(\s^1)^{ij}(\s^1)^{km} \right)(\g^\m)_{ab} \partial_\m \l_c^m   \cr
& + i (\cos 2 a_0 - \cos 2 b_0 )(\s^2)^{ij}(\s^2)^{km}\left((\g^5\g^\m)_{ab}(\g^5)_{c}^{~d}   \right. \cr
&\left. ~~~~~~~~~~~~~~+ C_{ab} (\g^\m)_{c}^{~d}+ (\g^5)_{ab}(\g^5\g^\m)_{c}^{~d}\right)\partial_\m \l_d^m   \cr
&+ \mbox{terms proportional to}~\sin 2 a_0~\mbox{and}~\sin 2 b_0~~~, } $$
%%%%%%%%%%%%%%%%%%%%%%%%%%%%%%%%%%%%%%%%%%%%
\be\eqalign{
\left\{ {\rm D}_a^i, {\rm D}_b^j \right\} {\rm d} = & 2 i \left( \d^{ij} + \sin 2 b_0~(\s^3)^{ij} \right) (\g^\m)_{ab} \partial_\m {\rm d}   \cr
&+2 i (\s^1)^{ij} \cos (a_0 + b_0) (\g^\m)_{ab} \partial_\m \tilde{\rm d}    \cr
& + 2 i (\s^2)^{ij} \sin(a_0 - b_0) (\g^5 \g^\n)_{ab} \partial^\m (\partial_\n \tilde{A}_\m - \partial_\m \tilde{A}_\n )~~~, \cr 
%%%%%%%%%%%%%%%%%%%%%%%%%%%%%%%%%%%%%%%%%%%%%%%
\left\{ {\rm D}_a^i, {\rm D}_b^j \right\} \tilde{\rm d} = & 2 i \left( \d^{ij} + \sin 2 a_0~(\s^3)^{ij} \right) (\g^\m)_{ab} \partial_\m \tilde{\rm d}   \cr
&+  2 i (\s^1)^{ij} \cos (a_0 + b_0) (\g^\m)_{ab} \partial_\m {\rm d}    \cr
& - 2 i (\s^2)^{ij} \sin(a_0 - b_0) (\g^5 \g^\n)_{ab} \partial^\m (\partial_\n A_\m - \partial_\m A_\n )~~~. 
}\ee
To have the canonical SUSY relationship to the momentum operator
on the right hand side for the bosons forces
\be\label{e:a0b0integer}
 a_0 = m \frac{\pi}{2}~~~,~~~b_0 = n \frac{\pi}{2}~~~,~~~m,n~~ \mbox{integers}
\ee
which makes
\be\label{e:abconstrained}\eqalign{
a^{ij} = & \cos m \frac{\pi}{2}~(\s^1)^{ij} + i \sin m \frac{\pi}{2}~(\s^2)^{ij}~~~, \cr
b^{ij} = & \cos n \frac{\pi}{2}~ \delta^{ij} + \sin n \frac{\pi}{2} ~ (\s^3)^{ij}~~~.
}\ee
Defining
\be\label{e:csdefs}\eqalign{
c_1 \equiv & \cos\left(\frac{(m+n)\pi}{2}\right)~~~, \cr
s_1 \equiv & \sin\left(\frac{(m-n)\pi}{2}\right)~~~, \cr
c_{2\pm} \equiv & \cos m\pi \pm \cos n\pi ~~~.
}\ee
the algebra becomes
\be\eqalign{
\left\{ {\rm D}_a^i, {\rm D}_b^j \right\} A_\nu = &  2 i \d^{ij} (\g^\m)_{ab} F_{\m\n}  + 2 i c_1\left(\s^1 \right)^{ij}  (\g^\m)_{ab} \tilde{F}_{\m\n}   \cr
&- i s_1 (\s^2)^{ij}  \e_\n^{~\m\a\b} (\g^5 \g_\m)_{ab} \tilde{F}_{\a\b} - 2 i s_1 (\s^2)^{ij}(\g^5\g_\n)_{ab} \tilde{\rm d}
~~~, \cr
%%%%%%%%%%%%%%%%%%%%%%%%%%%%%%%%%
\left\{ {\rm D}_a^i, {\rm D}_b^j \right\} \tilde{A}_\nu = &  2 i \d^{ij} (\g^\m)_{ab} \tilde{F}_{\m\n}  + 2 i c_1\left(\s^1 \right)^{ij} (\g^\m)_{ab} F_{\m\n}   \cr
&+ i s_1(\s^2)^{ij}  \e_\n^{~\m\a\b} (\g^5 \g_\m)_{ab} F_{\a\b} + 2 i s_1 (\s^2)^{ij} (\g^5\g_\n)_{ab} {\rm d}~~~, \cr
%%%%%%%%%%%%%%%%%%%%%%%%%%%%%%%%%%%%%%%%%%%%
\left\{ {\rm D}_a^i, {\rm D}_b^j \right\}  \l_c^k = & \left(2 i \d^{ij} \d^{km} + i c_{2+}(\s^1)^{ij}(\s^1)^{km} \right)(\g^\m)_{ab} \partial_\m \l_c^m   \cr
& + i c_{2-}(\s^2)^{ij}(\s^2)^{km}\left((\g^5\g^\m)_{ab}(\g^5)_{c}^{~d} + C_{ab} (\g^\m)_{c}^{~d}  \right. \cr
&\left. ~~~~~~~~~~~~~~~~~~~~~~~~~~~~~~~~+ (\g^5)_{ab}(\g^5\g^\m)_{c}^{~d}\right)\partial_\m \l_d^m
~~~, \cr
%%%%%%%%%%%%%%%%%%%%%%%%%%%%%%%%%%%%%%%%%%%%%%%
\left\{ {\rm D}_a^i, {\rm D}_b^j \right\} {\rm d} = & 2 i \d^{ij} (\g^\m)_{ab} \partial_\m {\rm d} + 2 i c_1 (\s^1)^{ij}  (\g^\m)_{ab} \partial_\m \tilde{\rm d}    \cr
& + 2 i  s_1 (\s^2)^{ij} (\g^5 \g^\n)_{ab} \partial^\m (\partial_\n \tilde{A}_\m - \partial_\m \tilde{A}_\n )~~~, \cr 
%%%%%%%%%%%%%%%%%%%%%%%%%%%%%%%%%%%%%%%%%%%%%%%
\left\{ {\rm D}_a^i, {\rm D}_b^j \right\} \tilde{\rm d} = & 2 i \d^{ij} (\g^\m)_{ab} \partial_\m \tilde{\rm d} + 2 i c_1 (\s^1)^{ij}  (\g^\m)_{ab} \partial_\m {\rm d}    \cr
& - 2 i s_1  (\s^2)^{ij}(\g^5 \g^\n)_{ab} \partial^\m (\partial_\n A_\m - \partial_\m A_\n )~~~. 
}\ee
We now dimensionally reduce to an eight by eight adinkra by considering all fields to have only temporal dependence. We choose the gauge
\be
  A_0 = \tilde{A}_0 = 0
\ee
and define
\be\eqalign{
    \lambda^1_1 &= i {\Psi}_1,~~~\lambda^1_2 = i{\Psi}_2,~~~\lambda^1_3 = i{\Psi}_3,~~~\lambda^1_4 = i {\Psi}_4, \cr
   \lambda^2_1 &= i {\Psi}_5,~~~\lambda^2_2 = i{\Psi}_6,~~~\lambda^2_3 = i{\Psi}_7,~~~\lambda^2_4 = i {\Psi}_8, \cr
    {\Phi}_1 &= A_1,~~~{\Phi}_2 = A_2,~~~{\Phi}_3 = A_3,~~~\partial_0{\Phi}_4 = {\rm d},\cr
   {\Phi}_5 &= \tilde{A}_1,~~~{\Phi}_6 = \tilde{A}_2,~~~{\Phi}_7 = \tilde{A}_3,~~~\partial_0{\Phi}_8 = \tilde{\rm d},
}\ee
and ${\rm D}_{\rm I}$ as before, Eq.~(\ref{e:Ddef}). The transformation laws reduce to the ever-now-more familiar form, Eq.~(\ref{eq:CCLR}), now with the L-matrices and R-matrices given in Appendix
D. These matrices satisfy the orthogonality relationship, Eq.~(\ref{e:orthogonal}), and the algebra
of (\ref{GarDNAlg3}) where 
\be\eqalign{
 \D{}_{{\rm I}{\rm J}}{}_i{}^k 
~=&~   2 c_1 \left(\sigma^1 \otimes \s^0 \otimes \s^0 \right){}_{{\rm I}{\rm J}} \left(\sigma^1\otimes \s^0 \otimes \s^0 \right)_i^{~k}   \cr
 &+ 2 s_1 \left(\sigma^2 \otimes \s^0 \otimes \s^2 \right){}_{{\rm I}{\rm J}} \left(\sigma^2\otimes \s^2 \otimes \s^3 \right)_i^{~k}   \cr
 &- 2 s_1 \left(\sigma^2 \otimes \s^2 \otimes \s^3 \right){}_{{\rm I}{\rm J}} \left(\sigma^2\otimes \s^0 \otimes \s^2 \right)_i^{~k} \cr
 &- 2 s_1 \left(\sigma^2 \otimes \s^2 \otimes \s^1 \right){}_{{\rm I}{\rm J}} \left(\sigma^2\otimes \s^2 \otimes \s^1 \right)_i^{~k}
~~~, }\ee
\noindent and
\be\eqalign{
{\Hat \D}{}_{{\rm I}{\rm J}}{}_\hi{}^\hk
~=&~  c_{2+} \left(\sigma^1 \otimes \s^0 \otimes \s^0 \right){}_{{\rm I}{\rm J}} \left(\sigma^1\otimes \s^0 \otimes \s^0 \right)_{\hat{i}}^{~\hat{k}}   \cr
 &+ c_{2-} \left(\sigma^2 \otimes \s^3 \otimes \s^2 \right){}_{{\rm I}{\rm J}} \left(\sigma^2\otimes \s^3 \otimes \s^2 \right)_{\hat{i}}^{~\hat{k}}   \cr
 &- c_{2-} \left(\sigma^2 \otimes \s^2 \otimes \s^0 \right){}_{{\rm I}{\rm J}} \left(\sigma^2\otimes \s^2 \otimes \s^0 \right)_{\hat{i}}^{~\hat{k}} \cr
 &+ c_{2-}\left(\sigma^2 \otimes \s^1 \otimes \s^2 \right){}_{{\rm I}{\rm J}} \left(\sigma^2\otimes \s^1 \otimes \s^2 \right)_{\hat{i}}^{~\hat{k}}~~~.
}\ee

The first order chromocharacters are as in Eq.~(\ref{e:ChromoOneCV}). For \emph{no} values of the integers $n$ and $m$ can the second order chromocharacters be made to satisfy Eq.~(\ref{e:ChromoTwoCV}). This is not a surprise as $n_c \ne n_t$ for this system. The second order chromocharacters take a complicated form, similar to that of the chiral-chiral system in Eq.~(\ref{e:CC2Chromos}), though even more complicated, and less enlightening. We have not completely worked out this formula, but we reiterate that we have proved by direct calculation that it \emph{can not} for any values of $n$ and $m$ take the form of Eq.~(\ref{e:ChromoTwoCV}).

 %%%%%%%%%%%%%%%%%%%%%%%%%%%%%%%%%%%%%%%%%%
 %%%%%%%%%%%%%%%%%%%%%%%%%%%%%%%%%%%%%%%%%%
 %%%%%%%%%%%%%%%%%%%%%%%%%%%%%%%%%%%%%%%%%%
\subsection{Building an $\bm {\cal N}$ $=$ 2 Supermultiplet From 
Tensor $+$ Tensor $\bm {\cal N}$ $=$ 1 Supermultiplets}\label{s:TT}

$~~~~$ For the case of two 4D, ${\cal N}$ $=$ 1 tensor multiplets, we introduce the 
transformation laws for this system as
\begin{equation}
    H_{\mu\nu\alpha} \equiv \partial_\mu B_{\n\a} +\partial_\a B_{\m\n}+\partial_\n B_{\a\m}~~~,~~~\tilde{H}_{\mu\nu\alpha} \equiv \partial_\mu \tilde{B}_{\n\a} +\partial_\a \tilde{B}_{\m\n}+\partial_\n \tilde{B}_{\a\m}
\end{equation}
the Lagrangian for this multiplet is
\begin{align}
   \mathcal{L} = &  - \frc{1}{3}H_{\mu\nu\alpha}H^{\mu\nu\alpha} - \frc{1}{3}\tilde{H}_{\mu\nu\alpha}\tilde{H}^{\mu\nu\alpha}- \frc{1}2 \partial_\mu \varphi \partial^\mu \varphi- \frc{1}2 \partial_\mu \tilde{\varphi} \partial^\mu \tilde{\varphi} + i \frc{1}2 (\g^\m)^{ab} \l_a^i \partial_\m \l_b^i 
\end{align}
which is an invariant of the transformation laws
\be\eqalign{
%%%%%%%%%%%%%%%%%%%%%%%%%%%%%%%%%%%%%%%%%%%%%%
{\rm D}_{a}^{i} \varphi = & b^{ij} \l_a^j~~~,\cr
%%%%%%%%%%%%%%%%%%%%%%%%%%%%%%%%%%%%%%%%%%%%%%
{\rm D}_{a}^{i} \tilde{\varphi} = & a^{ij} \l_a^j~~~,\cr
%%%%%%%%%%%%%%%%%%%%%%%%%%%%%%%%%%%%%%%%%%%%%%
{\rm D}_{a}^{i} B_{\mu\nu} =& - \frc{1}4 \left(\left[ \g_\m, \g_\n \right]\right)_a^{~b} b^{ij} \l_b^j ~~~,\cr
%%%%%%%%%%%%%%%%%%%%%%%%%%%%%%%%%%%%%%%%%%%%%%
{\rm D}_{a}^{i} \tilde{B}_{\mu\nu} =& - \frc{1}4 \left(\left[ \g_\m, \g_\n \right]\right)_a^{~b} a^{ij} \l_b^j ~~~,\cr
%%%%%%%%%%%%%%%%%%%%%%%%%%%%%%%%%%%%%%%%%%%%%%
{\rm D}_{a}^{i} \l_b^j =&  a^{ij}\left( i (\g^\m)_{ab} \partial_\m \tilde{\varphi} - (\g^5 \g_\m)_{ab} 
\e^{\m\rho\sigma\tau}\partial_\rho  \tilde{B}_{\sigma\tau} \right)   \cr
&+ b^{ij} \left( i (\g^\m)_{ab} \partial_\m \varphi - (\g^5 \g_\m)_{ab} \e^{\m\rho\sigma\tau} \partial_\rho B_{\sigma\tau} \right)
%%%%%%%%%%%%%%%%%%%%%%%%%%%%%%%%%%%%%%%%%%%%%%%%%%%%%%%%%%%%%%%
}\ee

For canonical bosonic momentum in the algebra, Eqs.~(\ref{e:a0b0integer}) and (\ref{e:abconstrained}) are once again forced on us and the algebra is
\be\eqalign{
%%%%%%%%%%%%%%%%%%%%%%%%%%%%%%%%%%%
\{ {\rm D}_a^i , {\rm D}_b^j \} \varphi =& 2 i \d^{ij} (\g^\m)_{ab} \partial_\m \varphi + 2 i c_1 (\s^1)^{ij} (\g^\m)_{ab}\partial_\m \tilde{\varphi}   \cr 
&- i \frc{2}3 s_1 (\s^2)^{ij}\e_{\m\n\a\b} (\g^5\g^\m)_{ab} \tilde{H}^{\n\a\b} ~~~, \cr
%%%%%%%%%%%%%%%%%%%%%%%%%%%%%%%%%%%%%
\{ {\rm D}_a^i , {\rm D}_b^j \} \tilde{\varphi}=& 2 i \d^{ij} (\g^\m)_{ab} \partial_\m \tilde{\varphi} + 2 i c_1 (\s^1)^{ij} (\g^\m)_{ab}\partial_\m \varphi   \cr 
&+i \frc{2}3 s_1 (\s^2)^{ij}\e_{\m\n\a\b} (\g^5\g^\m)_{ab} H^{\n\a\b} ~~~, \cr
%%%%%%%%%%%%%%%%%%%%%%%%%%%%%%%%%%%%%
\{ {\rm D}_a^i , {\rm D}_b^j \} B_{\m\n} = & 2 i \d^{ij} (\g^\a)_{ab} H_{\a\m\n} - i \d^{ij}(\g_{[\m})_{|ab|}\partial_{\n|} \varphi   \cr
&- i c_1 (\s^1)^{ij} (\g_{[\m})_{|ab|} \partial_{\n]} \tilde{\varphi} + i s_1 (\s^2)^{ij} \e_{\m\n}^{~~\a\b}(\g^5\g_\a)_{ab} \partial_\b \tilde{\varphi}   \cr 
&+ 2 i c_1 (\s^1)^{ij} (\g^\a)_{ab} \tilde{H}_{\a\m\n} +  i \frc{1}{3} s_1 (\s^2)^{ij} (\g^5\g_{[\m})_{|ab|}\e_{\n]\r\a\b}\tilde{H}^{\r\a\b}~~~, \cr
%%%%%%%%%%%%%%%%%%%%%%%%%%%%%%%%%%%%%
\{ {\rm D}_a^i , {\rm D}_b^j \} \tilde{B}_{\m\n} = & 2 i \d^{ij} (\g^\a)_{ab} \tilde{H}_{\a\m\n} - i \d^{ij}(\g_{[\m})_{|ab|}\partial_{\n]} \tilde{\varphi}   \cr
&- i c_1 (\s^1)^{ij} (\g_{[\m})_{|ab|} \partial_{\n]} \varphi - i s_1 (\s^2)^{ij} \e_{\m\n}^{~~\a\b}(\g^5\g_\a)_{ab} \partial_\b \varphi   \cr 
&+ 2 i c_1 (\s^1)^{ij} (\g^\a)_{ab} H_{\a\m\n} - i \frc{1}{3} s_1 (\s^2)^{ij} (\g^5\g_{[\m})_{|ab|}\e_{\n]\r\a\b}H^{\r\a\b}~~~, \cr
%%%%%%%%%%%%%%%%%%%%%%%%%%%%%%%%%%%%%%%%%%%%%%%%%%%%%%%%%%
\{ {\rm D}_a^i , {\rm D}_b^j \} \l_c^k = & i \left( 2 \d^{ij}\d^{kl} + c_{2+} (\s^1)^{ij}(\s^1)^{kl} \right)(\g^\m)_{ab} \partial_\m \l_c^l   \cr
&-i c_{2-} (\s^2)^{ij}(\s^2)^{kl} \left((\g^5\g^\m)_{ab}(\g^5)_{c}^{~d} + C_{ab} (\g^\m)_{c}^{~d} + \right. \cr
&\left. ~~~~~~~~~~~~~~~~~~~~~~~~~~~~~~~~- (\g^5)_{ab}(\g^5\g^\m)_{c}^{~d}\right)\partial_\m \l_d^l~~~,
}\ee
Choosing the gauge
\be
  B_{0i} = 0 = \tilde{B}_{0i}=0~~~,
\ee
defining
\be
  B_{0i} = 0 = \tilde{B}_{0i}=0~~~,
\ee
and
\be\eqalign{
{\Phi}_1 = \varphi ~~~,~~~{\Phi}_2 = 2 B_{12}~~~,&~~~{\Phi}_3 = 2 B_{23}~~~,~~~{\Phi}_4 = 2 B_{31}~~~, \cr
{\Phi}_5 = \tilde{\varphi} ~~~,~~~{\Phi}_2 = 2 \tilde{B}_{12}~~~,&~~~{\Phi}_3 = 2 \tilde{B}_{23}~~~,~~~{\Phi}_4 = 2 \tilde{B}_{31}~~~, \cr
i {\Psi}_1 = \l_1^1~~~,~~~i {\Psi}_2 = \l_2^1~~~&,~~~i {\Psi}_3 = \l_3^1~~~,~~~i {\Psi}_4 = \l_4^1~~~, \cr
i {\Psi}_5 = \l_1^2~~~,~~~i {\Psi}_6 = \l_2^2~~~&,~~~i {\Psi}_7 = \l_3^2~~~,~~~i {\Psi}_8 = \l_4^2~~~,
}\ee
and considering only temporal dependence of the fields reduces the transformation laws to Eq.~(\ref{eq:CCLR}) with the ${\rm D}_{\rm I}$ identifications as in Eq.~(\ref{e:Ddef}). The adinkra matrices in this basis are given in Appendix E and satisfy the algebra of Eq.~(\ref{GarDNAlg3}) with
\begin{align}
 \D{}_{{\rm I}{\rm J}}{}_i{}^k 
~=&~   2 c_1 \left(\sigma^1 \otimes \s^0 \otimes \s^0 \right){}_{{\rm I}{\rm J}} \left(\sigma^1\otimes \s^0 \otimes \s^0 \right)_i^{~k}    \nonumber\\*
 &- 2 s_1 \left(\sigma^2 \otimes \s^0 \otimes \s^2 \right){}_{{\rm I}{\rm J}} \left(\sigma^2\otimes \s^2 \otimes \s^1 \right)_i^{~k} 
~~~,  \nonumber\\*
~&~~   - 2 s_1 \left(\sigma^2 \otimes \s^2 \otimes \s^3 \right){}_{{\rm I}{\rm J}} \left(\sigma^2\otimes \s^0 \otimes \s^2 \right)_i^{~k} \nonumber\\*
 &- 2 s_1 \left(\sigma^2 \otimes \s^2 \otimes \s^1 \right){}_{{\rm I}{\rm J}} \left(\sigma^2\otimes \s^2 \otimes \s^3 \right)_i^{~k}
~~~, 
\end{align}
\noindent and
\be\eqalign{
{\Hat \D}{}_{{\rm I}{\rm J}}{}_\hi{}^\hk
~=&~  c_{2+} \left(\sigma^1 \otimes \s^0 \otimes \s^0 \right){}_{{\rm I}{\rm J}} \left(\sigma^1\otimes \s^0 \otimes \s^0 \right)_{\hat{i}}^{~\hat{k}}   \cr
 &- c_{2-} \left(\sigma^2 \otimes \s^3 \otimes \s^2 \right){}_{{\rm I}{\rm J}} \left(\sigma^2\otimes \s^3 \otimes \s^2 \right)_{\hat{i}}^{~\hat{k}}   \cr
 &- c_{2-} \left(\sigma^2 \otimes \s^2 \otimes \s^0 \right){}_{{\rm I}{\rm J}} \left(\sigma^2\otimes \s^2 \otimes \s^0 \right)_{\hat{i}}^{~\hat{k}} \cr
 &- c_{2-}\left(\sigma^2 \otimes \s^1 \otimes \s^2 \right){}_{{\rm I}{\rm J}} \left(\sigma^2\otimes \s^1 \otimes \s^2 \right)_{\hat{i}}^{~\hat{k}}~~~.
}\ee
The chromocharacters are \emph{exactly the same} as for the ${\mathcal N} = 2$ vector + vector multiplet. The most crucial result of these calculations is that for the tensor + tensor multiplet the second order chromocharacters \emph{can not} take the form of~Eq.~(\ref{e:ChromoTwoCV}) for any values of the integers $n$ and $m$. 

 %%%%%%%%%%%%%%%%%%%%%%%%%%%%%%%%%%%%%%%%%%
 %%%%%%%%%%%%%%%%%%%%%%%%%%%%%%%%%%%%%%%%%%
 %%%%%%%%%%%%%%%%%%%%%%%%%%%%%%%%%%%%%%%%%%
\subsection{Building an $\bm {\cal N}$ $=$ 2 Supermultiplet From 
Vector $+$ Tensor $\bm {\cal N}$ $=$ 1 Supermultiplets}\label{s:VT}

$~~~$ This supermultiplet has been discussed since the pioneering work of \cite{SSW}.  
The Lagrangian for this multiplet is
\begin{align}
   \mathcal{L} = & -\frc{1}4 F_{\mu\nu}F^{\mu\nu} - \frc{1}{3}H_{\mu\nu\alpha}H^{\mu\nu\alpha} - \frc{1}2 \partial_\mu \varphi \partial^\mu \varphi + i \frc{1}2 (\g^\m)^{ab} \l_a^i \partial_\m \l_b^i + \frc{1}2 {\rm d}^2
\end{align}
which is an invariant of the transformation laws
\be\eqalign{
%%%%%%%%%%%%%%%%%%%%%%%%%%%%%%%%%%%%%%%%%%%%%%
{\rm D}_{a}^{i} \varphi = & b^{ij} \l_a^j~~~,\cr
%%%%%%%%%%%%%%%%%%%%%%%%%%%%%%%%%%%%%%%%%%%%%%
{\rm D}_{a}^{i} A_\mu = &(\g_\m)_a^{~b} a^{ij} \l_b^j ~~~,\cr
%%%%%%%%%%%%%%%%%%%%%%%%%%%%%%%%%%%%%%%%%%%%%%
{\rm D}_{a}^{i} B_{\mu\nu} =& - \frc{1}4 \left(\left[ \g_\m, \g_\n \right]\right)_a^{~b} b^{ij} \l_b^j ~~~,\cr
%%%%%%%%%%%%%%%%%%%%%%%%%%%%%%%%%%%%%%%%%%%%%%
{\rm D}_{a}^{i} \l_b^j =&  a^{ij} \left( - \frc{i}4\left(\left[ \g^\m , \g^\n \right]\right)_{ab} F_{\mu\nu}   + (\g^5)_{ab} {\rm d} \right) \cr
&+ b^{ij} \left( i (\g^\m)_{ab} \partial_\m \varphi - (\g^5 \g_\m)_{ab} \e^{\m\rho\sigma\tau}\partial_\rho B_{\sigma\tau} \right)
~~~, \cr
%%%%%%%%%%%%%%%%%%%%%%%%%%%%%%%%%%%%%%%%%%%%%%%%%%%%%%%
{\rm D}_a^i {\rm d} = & i ( \g^5 \g^\m)_{a}^{~b} a^{ij} \partial_\m \l_b^j
%%%%%%%%%%%%%%%%%%%%%%%%%%%%%%%%%%%%%%%%%%%%%%%%%%%%%%%%%%%%%%%
}\ee
For the canonical bosonic momentum term to appear in the algebra, Eqs.~(\ref{e:a0b0integer}) and (\ref{e:abconstrained}) are once again forced on us and the algebra is 
\begin{align*}
%%%%%%%%%%%%%%%%%%%%%%%%%%%%%%%%%%%
\{ {\rm D}_a^i , {\rm D}_b^j \} \varphi =& 2 i \d^{ij} (\g^\m)_{ab} \partial_\m \varphi + 2 i s_1 (\s^2)^{ij} (\g^5)_{ab}\partial_\m {\rm d}   \nonumber\\*
&- i c_1 (\s^1)^{ij}(\g^\m\g^\n)_{ab}F_{\m\n}~~~, \nonumber\\*
%%%%%%%%%%%%%%%%%%%%%%%%%%%%%%%%%%%%%
\left\{ {\rm D}_a^i, {\rm D}_b^j \right\} A_\nu = &  2 i \d^{ij} (\g^\m)_{ab} F_{\m\n}  + 2 i c_1 \left(\s^1 \right)^{ij} (\g^\a\g^\m)_{ab} H_{\a\m\n}   \nonumber\\*
&- i\frc{2}3 s_1 (\s^2)^{ij} (\g^5)_{ab}\e_{\n\a\b\m} H^{\a\b\m} + i c_1 (\s^1)^{ij} \left(\left[ \g_\n, \g^\m \right]\right)_{ab}\partial_\m\varphi  ~~~,  {~~~~~~~~~~~~~~~}
\nonumber\\*
%%%%%%%%%%%%%%%%%%%%%%%%%%%%%%%%%%%
&- 2 s_1 C_{ab}(\s^2)^{ij}\partial_\n\varphi ~~~,
\end{align*}
\be\eqalign{
%%%%%%%%%%%%%%%%%%%%%%%%%%%%%%%%%%%%%%%%%%%%
\{ {\rm D}_a^i , {\rm D}_b^j \} B_{\m\n} = & 2 i \d^{ij} (\g^\a)_{ab} H_{\a\m\n} - s_1 (\s^2)^{ij}C_{ab} F_{\m\n}   \cr 
&+ i \frc{1}2 s_1  (\s^2)^{ij} (\g^5)_{ab}\e_{\m\n\a\b} F^{\a\b} + i \frc{1}2 c_1   (\s^1)^{ij} \left(\left[ \g_{[\m} , \g^\a \right]\right)_{|ab|} F_{\n ] \a}   \cr
&+ \frc{1}2 c_1 (\s^1)^{ij} \left( \g^5 \left[ \g_\m , \g_\n \right] \right)_{ab} {\rm d} - i \d^{ij} (\g_{[\m})_{|ab|} \partial_{\n]}\varphi ~~~,\cr
%%%%%%%%%%%%%%%%%%%%%%%%%%%%%%%%%%%%%%%%%%%%%%%%%%%%%%%%%%
\{ {\rm D}_a^i , {\rm D}_b^j \} \l_c^k = & i \left( 2 \d^{ij}\d^{kl} + c_{2+} (\s^1)^{ij}(\s^1)^{kl} \right)(\g^\m)_{ab} \partial_\m \l_c^l   \cr
&+i c_{2+} (\s^2)^{ij}(\s^2)^{kl} \left((\g^5\g^\m)_{ab}(\g^5)_{c}^{~d} + C_{ab} (\g^\m)_{c}^{~d} \right)\partial_\m \l_d^l   \cr
& +i c_{2-} (\s^2)^{ij}(\s^2)^{kl}(\g^5)_{ab}(\g^5\g^\m)_{c}^{~d} \partial_\m \l_d^l~~~, \cr
%%%%%%%%%%%%%%%%%%%%%%%%%%%%%%%%%%%%%%%%%%%%%%%
\left\{ {\rm D}_a^i, {\rm D}_b^j \right\} {\rm d} = & 2 i \d^{ij} (\g^\m)_{ab} \partial_\m {\rm d} - 2 c_1 (\s^1)^{ij} (\g^5\g^\m\g^\n)_{ab}\partial^\a H_{\a\m\n}   \cr
&- 2 i s_1 (\s^2)^{ij} (\g^5)_{ab} \partial_\m \partial^\m \varphi ~~~.
}\ee

Choosing the gauge
\be
  B_{0i} = A_0 = 0~~~,
\ee
defining
\be\eqalign{
{\Phi}_1 = \varphi ~~~,~~~{\Phi}_2 = 2 B_{12}~~~,&~~~{\Phi}_3 = 2 B_{23}~~~,~~~{\Phi}_4 = 2 B_{31}~~~, \cr
{\Phi}_5 = A_1 ~~~,~~~{\Phi}_2 = A_2~~~,&~~~{\Phi}_3 =A_3~~~,~~~{\Phi}_4 =  \int dt~{\rm d} ~~~, \cr
i {\Psi}_1 = \l_1^1~~~,~~~i {\Psi}_2 = \l_2^1~~~&,~~~i {\Psi}_3 = \l_3^1~~~,~~~i {\Psi}_4 = \l_4^1~~~, \cr
i {\Psi}_5 = \l_1^2~~~,~~~i {\Psi}_6 = \l_2^2~~~&,~~~i {\Psi}_7 = \l_3^2~~~,~~~i {\Psi}_8 = \l_4^2~~~,
}\ee
and considering only temporal dependence of the fields reduces the transformation laws to Eq.~(\ref{eq:CCLR}) with the ${\rm D}_{\rm I}$ identifications as in Eq.~(\ref{e:Ddef}). The adinkra matrices in this basis are given in Appendix F. They satisfy the orthogonality relationship, Eq.~(\ref{e:orthogonal}), and the algebra of Eq.~(\ref{GarDNAlg3}) with $\Delta_{{\rm IJ}i}^{~~~k}$ and ${\Hat \D}{}_{{\rm I}{\rm J}}{}_\hi{}^\hk$ given  by
\be\eqalign{
 \D{}_{{\rm I}{\rm J}}{}_i{}^k 
~=&~   2 c_1 \left(\sigma^1 \otimes \s^0 \otimes \s^3 \right){}_{{\rm I}{\rm J}} (\Delta^{(VT)}_1 ){}_i^{~k}   \cr
 &- 2 c_1 \left(\sigma^1 \otimes \s^0 \otimes \s^1 \right){}_{{\rm I}{\rm J}} (\Delta^{(VT)}_2){}_i^{~k}   \cr
 &- 2 c_1 \left(\sigma^1 \otimes \s^2 \otimes \s^2 \right){}_{{\rm I}{\rm J}} (\Delta^{(VT)}_3){}_i^{~k} \cr
 &- 2 s_1 \left(\sigma^2 \otimes \s^2 \otimes \s^0 \right){}_{{\rm I}{\rm J}} (\Delta^{(VT)}_4){}_i^{~k}
~~~, }\ee
\noindent where the quantities $\Delta^{(VT)}_1$, etc. are defined by
$$  \eqalign{
 \Delta^{(VT)}_1  ~=~ \left[\begin{array}{cc}
 0  &   (8)_b    (1324)  \\
(1)_b (1423)    &0    \\
\end{array}\right]   ~~~~~~,~~ }
$$
\be  \eqalign{
\Delta^{(VT)}_2  ~=~  \left[\begin{array}{cc}
0  &   (13)_b    (34)  \\
(13)_b    (34)    &  0    \\
\end{array}\right] 
~~~~~\,~~~, ~\,~  \cr
 \Delta^{(VT)}_3  ~=~  \left[\begin{array}{cc}
 0  &   (11)_b    (12)  \\
 (11)_b    (12)  &  0    \\
\end{array}\right] 
   ~~~~\,~\,~\,\,~,\,~~    \cr
 \Delta^{(VT)}_4  ~=~  \left[\begin{array}{cc}
 0  &   (1)_b    (1423)  \\
 (8)_b    (1324)  &  0    \\
\end{array}\right] 
   ~~~~~\,~,~~    }
\ee
\noindent where we have used the Boolean Factor notation of \cite{permutadnk} to indicate locations of minus signs for permutation matrices defined as, for instance: 
\begin{align}
	 (8)_b    (1324) \equiv \left( 
	 \begin{array}{cccc}
	    0 & 0 & 1 & 0\\
	    0 & 0 & 0 & 1\\
	    0 & 1 & 0 & 0 \\
	    -1 & 0 & 0 & 0
	 \end{array}
	 \right)~~~,~~~(11)_b    (12) \equiv \left( 
	 \begin{array}{cccc}
	    0 & -1 & 0 & 0\\
	    -1 & 0 & 0 & 0\\
	    0 & 0 & 1 & 0 \\
	    0 & 0 & 0 & -1
	 \end{array}
	 \right)~~~.
\end{align}
We also have
\be\eqalign{
{\Hat \D}{}_{{\rm I}{\rm J}}{}_\hi{}^\hk
~=&~  c_{2+} \left(\sigma^1 \otimes \s^0 \otimes \s^0 \right){}_{{\rm I}{\rm J}} \left(\sigma^1\otimes \s^0 \otimes \s^0 \right)_{\hat{i}}^{~\hat{k}}   \cr
 &+ c_{2+} \left(\sigma^2 \otimes \s^3 \otimes \s^2 \right){}_{{\rm I}{\rm J}} \left(\sigma^2\otimes \s^3 \otimes \s^2 \right)_{\hat{i}}^{~\hat{k}}   \cr
 &+ c_{2+} \left(\sigma^2 \otimes \s^1 \otimes \s^2 \right){}_{{\rm I}{\rm J}} \left(\sigma^2\otimes \s^1 \otimes \s^2 \right)_{\hat{i}}^{~\hat{k}} \cr
 &- c_{2-}\left(\sigma^2 \otimes \s^2 \otimes \s^0 \right){}_{{\rm I}{\rm J}} \left(\sigma^2\otimes \s^2 \otimes \s^0 \right)_{\hat{i}}^{~\hat{k}}~~~.
}\ee

The first order chromocharacters satisfy Eq.~(\ref{e:ChromoOneCV}). The second order chromocharacters \emph{can not} take the form of~Eq.~(\ref{e:ChromoTwoCV}) for any values of the integers $n$ and $m$. In summary, none of the ${\mathcal N} = 2$ supermultiplets from the list of chiral + chiral, vector + vector, and tensor + tensor, nor vector+tensor have second order chromocharacters as given in Eq.~(\ref{e:ChromoTwoCV}).

$~~~~$ 
\newpage

\subsection{Summary of Building ${\cal N} = 2$ Multiplets from ${\cal N} = 1$ Multiplets}
$~~~~$ To recapitulate the results of this chapter, we have seen by explicit
construction that there are six possible pairings of 4D, $\cal N$ $=$ 
1 chiral, vector, and tensor multiplets that may be taken as starting 
points in an attempt to construct 4D, $\cal N$ $=$ 2 supermultiplets 
that are:  \newline \indent
(a.) completely off-shell (i.\ e.\  require no a priori differential 
constraints  \newline \indent $~~~~~$
 imposed on any fields), and  \newline \indent
(b.) require no off-shell central charges. \newline \noindent
However, the result of this study is that only two combinations:
 \newline \indent
(a.) chiral $+$ vector, and  \newline \indent
(b.)  chiral $+$ tensor,  \newline \noindent
satisfy the required conditions stated immediately above.

The following two questions seem important to ask. ``Why do the results work 
out in this way?''   ``What is it that distinguishes two of the six possible starting 
points from the others?''   Simply reporting these results does nothing to reveal 
what deeper mathematical structures impose these results.  

If we use only the conventional and traditional approaches to analyzing these
results, there is no simple and elegant way (at least known to these authors) to 
answer these questions.  The situation is vaguely analogous to looking at the 
quark model and asking, ``Which composite systems of quarks occur as an 
observable baryons?''  The answer is well known, ``All observable baryonic 
composite states must have vanishing color quantum number.''  In the next 
section, we will argue that the adinkra-based model of off-shell SUSY 
representations provides a remarkably elegant and simple answer to the 
questions above and does so in a manner similar to the confinement of QCD 
color.

 %%%%%%%%%%%%%%%%%%%%%%%%%%%%%%%%%%%%%%%%%%
 %%%%%%%%%%%%%%%%%%%%%%%%%%%%%%%%%%%%%%%%%%
 %%%%%%%%%%%%%%%%%%%%%%%%%%%%%%%%%%%%%%%%%%
\section{Adinkra `Color-like' Confinement Rules For 4D, $\bm {\cal N}$ $=$ 1 Reps
Within Off-Shell $\bm {\cal N}$ $=$ 2 Supermultiplets }

$~~~~$ The survey of building ${\mathcal N} = 2$ supermultiplets from ${\mathcal N} = 1$
supermultiplets shows there appears to be a `super-
selection-like rule' that governs the ${\mathcal N} = 1$ content of 
the ${\mathcal N} = 2$ extended supermultiplets.

In the work of \cite{G-1}, it was argued that the chromocharacters associated 4D, 
$\cal N$ $=$ 1 supermultiplets generally take the form
\be  \eqalign{
\varphi^{(2)} {}_{\bj I }  {}_{\bj J }  {}_{\bj K }  {}_{\bj L } ({\cal N} \, =\, 1)
~&=~   4 \, \left( \, n_c \,+\, n_t  \, \right)  \left[~ \delta_{ {}_{\bj I }  {}_{\bj J }}\delta_{ 
{}_{\bj K }  {}_{\bj L }} ~-~  \delta_{ {}_{\bj I }  {}_{\bj K }}
\delta_{ {}_{\bj J }  {}_{\bj L }} ~+~ \delta_{ {}_{\bj I }  {}_{
\bj L }}\delta_{ {}_{\bj J }  {}_{\bj K }} ~ \right] ~+~ 4 \, \chi_{\rm o} \, 
\epsilon_{ {}_{\bj I }  {}_{\bj J }  {}_{\bj K }  {}_{\bj L }}  ~~,  \cr
\chi_{\rm o}~&=~    \, \left( \, n_c \,-\, n_t  \, \right)  \, 
 ~~~, \cr
}
\label{Chr0M3}
\ee
and we notice the appearance of the Levi-Civita tensor is allowed because 
4D, $\cal N$ $=$ 1 supersymmetric multiplets have only four supercharges
and thus require only $O(4)$ symmetry of their chromocharacters.  It follows 
that the quantity $\chi_{\rm o}$ (`Kye-Oh') can be abstracted from
\be  \eqalign{
\chi_{\rm o} ~=~ \fracm 1{\,4\, \cdot \, 4!} \, \epsilon^{ {}^{\bj I }  {}^{\bj J }  {}^{\bj K }  
 {}^{\bj L }} \, \varphi^{(2)} {}_{\bj I }  {}_{\bj J }  {}_{\bj K }  {}_{\bj L } ({\cal N} 
 \, =\, 1)   ~~~.
 }
\label{Chr0M3a}
\ee

In every system investigated in chapter three, if one begins with off-shell $ {\cal N}$ $=$ 1 
supermultiplets and uses these as a basis for constructing  off-shell $ {\cal N}$ $=$ 2 
supermultiplets, the latter will not be off-shell and free of central charges unless 
$ \S \left( n_c  ~-~   n_t \right) $ $=$ 0, where the sum is taken over the $\cal N$ 
$=$ 1 supermultiplets.

This observation is a very explicit demonstration of the utility of the adinkra-based
view that has been developed in a number of our past works.  Taking the 
adinkra approach \cite{GRana} - \cite{Bowtie}, one is naturally led to the
existence of $n_c$, and $n_t$.  Below we will give a simple explanation 
on why this super-selection-like rule must appear in all supermultiplets that arise in 
the adinkra approach.  In the process, we will show that the adinkra-based approach
thus leads to a new and effective tool, which is obscured in more conventional 
approaches, for understanding fundamental aspects of SUSY representation 
theory in four dimensions.

Let us consider the chromocharacter in (\ref{Cchrm}) for the case of $p$ $=$ 2.  
All our previous works suggests that the chromocharacters possess $SO(N)$ symmetry,
i.\ e.\ $SO(8)$ symmetry for our considerations.  This is a very powerful assertion and 
we will now argue that it is the cause of the proposed super-selection-like rule.
Due to $SO(8)$ symmetry, the form of the second order chromocharacter in this case must be
\be \eqalign{   {~~~~~~}
\varphi^{(2)} {}_{{\bj I}_{{}_1} } {}_{{\bj J}_{{}_1} }  {}_{{\bj I}_{{}_2} } {}_{{\bj J}_{{}_2} }
({\cal N} \, =\, 2)
~\propto ~\left[ \,  \d{}_{{\bj I}_{{}_1} } {}_{{\bj J}_{{}_1} }    \d{}_{{\bj I}_{{}_2} } {}_{{\bj 
J}_{{}_2} }  ~-~ \d{}_{{\bj I}_{{}_1} } {}_{{\bj I}_{{}_2} }    \d{}_{{\bj J}_{{}_1} } {}_{{\bj 
J}_{{}_2} }   ~+~ \d{}_{{\bj I}_{{}_1} } {}_{{\bj J}_{{}_2} }    \d{}_{{\bj I}_{{}_2} } {}_{{\bj 
J}_{{}_1} }  \, \right] ~~~~~,
}
\label{eq:E23}
\ee
where we have used the properties of the $ (\,{\rm L}_{\rm I}\,)$-matrices to arrive at
this conclusion. This must be true for the chromocharacters associated with 
supermultiplets that possess 4D, $\cal N$ $=$ 2 supersymmetry by our SUSY
holography conjecture.

For the $\cal N$ $=$ 2 chromocharacter, the indices ${\rm I}_1$, ${\rm I}_2$,
${\rm J}_1$, and ${\rm J}_2$ take on values 1, $\dots$, 8 while for the 
$\cal N$ $=$ 1 chromocharacter, the indices ${\rm I}$, ${\rm J}$,
${\rm K}$, and ${\rm L}$ take on values 1, $\dots$, 4.  So in order to
compare these two chromocharacter formulae, one must perform a
projection of $SO(8)$ down to $SO(4)$.  However, such a projection will
{\em {never}} create a term proportional to the Levi-Civita tensor.

Thus, these considerations lead to the conclusion that for all 4D, $\cal N$ $=$ 1
supermultiplets that occur as sub-supermultiplets within off-shell 4D, $\cal N$ $=$ 2
supermultiplets, the value of $\chi_{\rm o}$ when summed over the
$\cal N$ $=$ 1 sub-supermultiplets must vanish.  It is very satisfying to see
that this formal argument is in agreement with the explicit calculations
performed in the previous chapters.

  %%%%%%%%%%%%%%%%%%%%%%%%%%%%%%%%%%%%%%%%%%
 %%%%%%%%%%%%%%%%%%%%%%%%%%%%%%%%%%%%%%%%%%
 %%%%%%%%%%%%%%%%%%%%%%%%%%%%%%%%%%%%%%%%%%
\section{Seeing `Kye-Oh' in 4D, $\bm {\cal N}$ $=$ 1 
Supermultiplets \\ Without 0-Brane Reduction}

$~~~~$ In this chapter, we will make an observation about the determination of 
 $\chi_{\rm o}$ that shows its value on these three supermultiplets can be
found {\em {without}} actually carrying out 0-brane reduction.   We find this
is an interesting result as it will show that $\chi_{\rm o}$ can be directly determined
by a calculation in four dimensions.

In this chapter, we are going to use the conventions of {\em {Superspace}} where
two-component Weyl spinors have been our tradition.  To facilitate this, we first
establish a dictionary between the conventions of \cite{G-1} and {\em {Superspace}}
\cite{SUSYBk}.  Using the former we have
\be \eqalign{
&~~~~~~~~~~~~~~~~~~~ {(\gamma^0)}{}_a{}^b  = i ( \sigma^3
 \otimes \sigma^2  ){}_a{}^b 
~~~~,~~~~~~ {(\gamma^1)}{}_a{}^b  = ({\bf I}_2 
\otimes \sigma^1 ){}_a{}^b ~~~~~, \cr
&~~~~~~~~~~~~~~~~~~~ {(\gamma^2)}{}_a{}^b  = (\sigma^2 
\otimes \sigma^2 ){}_a{}^b ~~~~~,~~~~~~ 
  {(\gamma^3)}{}_a{}^b  = ({\bf I}_2 
\otimes \sigma^3  ){}_a{}^b  ~~~~~,
\cr
&~~~~~~~~~~~~~~~~~~~~~~~~~~~~~~~~~~~
 {(\gamma^5)}{}_a{}^b  = -(\sigma^1 \otimes \sigma^2 ){}_a{}^b ~~~~~,
\cr
&~~~~~~~~~~~~~~~  C_{ab} \equiv -i (\sigma^3 \otimes \sigma^2)_{a b} 
~=~ \left(\begin{array}{cccc} 0 & -1 & 0 & 0\\ 1 & 0 & 0 & 0\\ 0 & 0 & 0 & 1\\ 
0 & 0 & -1 & 0\end{array}\right)~~~\to ~~ C_{ab} = -C_{ba} ~~~.
} \label{KhyO1}
\ee
The inverse spinor metric is defined by the condition $C^{ab}C_{ac} = \delta{}_c{}^b$.

The chiral projection operators $({\bf P}_{\pm})$ are defined by 
\be
({\bf P}_{\pm}){}_a{}^b ~=~ \fracm 12 \, \big[ ~   ({\bf I}_4){}_a{}^b ~\pm~  {(\gamma^5)}{}_a{}^b ~ \big]
 \label{KhyO2}
\ee
which implies
\be \eqalign{
({\bf P}_{+}){}_a{}^b {\rm D}{}_b ~=~ \fracm 12 \,  \left[\begin{array}{c}
~\left( \, {\rm D}_{1} \, + \, i \, {\rm D}_{4}  \, \right) \\
~\left( \,  {\rm D}_{2} \, - \, i \, {\rm D}_{3}  \, \right) \\
~\left( \,  {\rm D}_{3} \, + \, i \, {\rm D}_{2}  \, \right) \\
~  \left( \, {\rm D}_{4} \, - \, i \, {\rm D}_{1}   \, \right) \\
\end{array}\right]  ~=~ \fracm 12 \,  \left[\begin{array}{c}
~\left( \, {\rm D}_{1} \, + \, i \, {\rm D}_{4}  \, \right) \\
~\left( \,  {\rm D}_{2} \, - \, i \, {\rm D}_{3}  \, \right) \\
~i \, \left( \,  {\rm D}_{2} \, - \, i \, {\rm D}_{3}  \, \right) \\
~ - i \,  \left( \, {\rm D}_{1} \, + \, i \, {\rm D}_{4}   \, \right) \\
\end{array}\right]   ~~~,}
 \label{KhyO3}
\ee
\be \eqalign{
({\bf P}_{-}){}_a{}^b {\rm D}{}_b ~=~ \fracm 12 \,  \left[\begin{array}{c}
~\left( \, {\rm D}_{1} \, - \, i \, {\rm D}_{4}  \, \right) \\
~\left( \,  {\rm D}_{2} \, + \, i \, {\rm D}_{3}  \, \right) \\
~\left( \,  {\rm D}_{3} \, - \, i \, {\rm D}_{2}  \, \right) \\
~  \left( \, {\rm D}_{4} \, + \, i \, {\rm D}_{1}   \, \right) \\
\end{array}\right]  
%%%%%%%%%%%%%%%%%%%%%%%%%%%%%%
~=~ \fracm 12 \,  \left[\begin{array}{c}
~\left( \, {\rm D}_{1} \, - \, i \, {\rm D}_{4}  \, \right) \\
~\left( \,  {\rm D}_{2} \, + \, i \, {\rm D}_{3}  \, \right) \\
~- i \, \left( \,  {\rm D}_{2} \, + \, i \, {\rm D}_{3}  \, \right) \\
~ i \,  \left( \, {\rm D}_{1} \, - \, i \, {\rm D}_{4}   \, \right) \\
\end{array}\right]   ~~~,}
 \label{KhyO4}
\ee

Let us define
\be \eqalign{
D{}_{A} ~&=~ \fracm 1{\sqrt 2}  \, \left( \, {\rm D}_{1} \, + \, i \, {\rm D}_{4}  \, \right) 
~~~~~~~,~~~
D{}_{B}  ~=~ \fracm 1{\sqrt 2}  \, \left( \, {\rm D}_{2} \, - \, i \, {\rm D}_{3}  \, \right)  ~~~~\,~~~,
\cr
{\Bar D}{}_{\dot A} ~&=~- \,  \fracm 1{\sqrt 2}  \, \left( \, {\rm D}_{1} \, - \, i \, {\rm D}_{4}  \, \right) 
~~~,~~~
{\Bar D}{}_{\dot B} ~=~- \, \fracm 1{\sqrt 2}  \, \left( \, {\rm D}_{2} \, + \, i \, {\rm D}_{3}  \, \right)
~~~, 
}   \label{KhyO5}
\ee
here the subscripts $A$, $B$, $\dot A$ and $\dot B$ are understood 
to be labels, not indices taking on multiple values.  We next derive the form 
of the super algebra generated by these four spinorial derivative operators.  
We find
\begin{align}
{\big \{} \,   D{}_{A} ~,~ D{}_{A} \, {\big\}} ~&=~  \fracm 12 \, {\big[} \, {\big \{} \,  
{\rm D}_{1} ~,~ {\rm D}_{1}  \, {\big \}}  ~-~ {\big \{} \,  {\rm D}_{4} ~,~ {\rm D}_{4}  
\, {\big \}} ~+~ i 2\, {\big \{} \,  {\rm D}_{1} ~,~ {\rm D}_{4}  \, {\big \}} \, {\big ]}
~~~, \cr
%%%%%%%%%%%%%%%%%%%%%%%%%%%%%
~&=~ i  \, {\big[} \, (\g^{\mu})_{1 \, 1}  \pa_{\mu} ~-~ (\g^{\mu})_{4 \, 4}  
\pa_{\mu} \, {\big ]} ~-~ 2 \, (\g^{\mu})_{1 \, 4}  
\pa_{\mu}  ~~~, \cr
&{~}  \cr
%%%%%%%%%%%%%%%%%%%%%%%%%%%%%
%%%%%%%%%%%%%%%%%%%%%%%%%%%%%
{\big \{} \,   D{}_{A} ~,~ D{}_{B} \, {\big\}} ~&=~  \fracm 12 \, {\big[} \, {\big \{} \,  
{\rm D}_{1} ~,~ {\rm D}_{2}  \, {\big \}}  ~+~ {\big \{} \,  {\rm D}_{3} ~,~ {\rm D
}_{4}  \, {\big \}} ~+~ i \, {\big \{} \,  {\rm D}_{2} ~,~ {\rm D}_{4}  \, {\big \}}
~-~ i  {\big \{} \,  {\rm D}_{1} ~,~ {\rm D}_{3}  \, {\big \}} \, {\big ]}
~~~, \cr
%%%%%%%%%%%%%%%%%%%%%%%%%%%%%
~&=~ i  \, {\big[} \, (\g^{\mu})_{1 \, 2}  \pa_{\mu} ~+~ (\g^{\mu})_{3 \, 4}  
\pa_{\mu} \, {\big ]} ~-~  \, (\g^{\mu})_{2 \, 4}  
\pa_{\mu} ~+~  \, (\g^{\mu})_{1 \, 3}  
\pa_{\mu}  ~~~, \cr
&{~}  \cr
%%%%%%%%%%%%%%%%%%%%%%%%%%%%%
%%%%%%%%%%%%%%%%%%%%%%%%%%%%%
{\big \{} \,   D{}_{B} ~,~ D{}_{B} \, {\big\}} ~&=~   \fracm 12 \, {\big[} \, {\big \{} \,
{\rm D}_{2} ~,~ {\rm D}_{2}  \, {\big \}}  ~-~ {\big \{} \,  {\rm D}_{3} ~,~ {\rm D
}_{3}  \, {\big \}} ~-~ i 2\, {\big \{} \,  {\rm D}_{2} ~,~ {\rm D}_{3}  \, {\big \}} 
\, {\big ]} ~~~, \cr
%%%%%%%%%%%%%%%%%%%%%%%%%%%%%
~&=~ i  \, {\big[} \, (\g^{\mu})_{2 \, 2}  \pa_{\mu} ~-~ (\g^{\mu})_{3 \, 3}  
\pa_{\mu} \, {\big ]} ~+~ 2 \, (\g^{\mu})_{2 \, 3}  
\pa_{\mu}  ~~~, \cr
&{~}  \cr
%%%%%%%%%%%%%%%%%%%%%%%%%%%%%
%%%%%%%%%%%%%%%%%%%%%%%%%%%%%
{\big \{} \,   D{}_{A} ~,~ {\Bar D}{}_{\dot B} \, {\big\}} ~&=~ -\,  \fracm 12 \, {\big[} \, 
{\big \{} \,  {\rm D}_{1} ~,~ {\rm D}_{2}  \, {\big \}}  ~-~ {\big \{} \,  {\rm D}_{3} 
~,~ {\rm D}_{4}  \, {\big \}} ~+~ i \, {\big \{} \,  {\rm D}_{2} ~,~ {\rm D}_{4} \, 
{\big \}} ~+~ i \, {\big \{} \,  {\rm D}_{1} ~,~ {\rm D}_{3}  \, {\big \}} \, {\big ]}
~~~, \cr
%%%%%%%%%%%%%%%%%%%%%%%%%%%%%
~&=~-\,  i  \, {\big[} \, (\g^{\mu})_{1 \, 2}  \pa_{\mu} ~-~ (\g^{\mu})_{3 \, 4}  
\pa_{\mu} \, {\big ]} ~-~  \, (\g^{\mu})_{2 \, 4}  \pa_{\mu} ~-~  \, (\g^{
\mu})_{1 \, 3}  \pa_{\mu}  ~~~, \cr
&{~}  \cr
%%%%%%%%%%%%%%%%%%%%%%%%%%%%%
%%%%%%%%%%%%%%%%%%%%%%%%%%%%%
{\big \{} \,   D{}_{A} ~,~ {\Bar D}{}_{\dot A} \, {\big\}} ~&=~-\, \fracm 12 \, {\big[} \, 
{\big \{} \,  {\rm D}_{1} ~,~ {\rm D}_{1}  \, {\big \}}  ~+~ {\big \{} \,  {\rm D}_{4} 
~,~ {\rm D}_{4}  \, {\big \}} \, {\big ]}  \cr
%%%%%%%%%%%%%%%%%%%%%%%%%%%%%
~&=~-\,  i  \, {\big[} \, (\g^{\mu})_{1 \, 1}  \pa_{\mu} ~+~ (\g^{\mu})_{4 \, 4}  
\pa_{\mu} \, {\big ]} ~~~, \cr
&{~}  \cr
%%%%%%%%%%%%%%%%%%%%%%%%%%%%%
%%%%%%%%%%%%%%%%%%%%%%%%%%%%%
{\big \{} \,   D{}_{B} ~,~ {\Bar D}{}_{\dot B} \, {\big\}} ~&=~ -\, \fracm 12 \, {\big[} \, 
{\big \{} \,  {\rm D}_{2} ~,~ {\rm D}_{2}  \, {\big \}}  ~+~ {\big \{} \,  {\rm D}_{3} 
~,~ {\rm D}_{3}  \, {\big \}}  \, {\big ]}  \cr
%%%%%%%%%%%%%%%%%%%%%%%%%%%%%
~&=~-\,  i  \, {\big[} \, (\g^{\mu})_{2 \, 2}  \pa_{\mu} ~+~ (\g^{\mu})_{3 \, 3}  \pa_{\mu} \, {\big ]} 
 \label{KhyO6}
\end{align}
It is a straightforward exercise to show that given the representation
for the gamma matrices we use further imply
\be \eqalign{
& {(\gamma^0)}{}_a{}_b  =  ( {\bf I}_2
 \otimes {\bf I}_2  ){}_a{}_b
 = \left[\begin{array}{cccc} 
  1 & 0 & 0 & 0\\ 
 %%%%%%%%
 0 & \, 1 & 0 & 0\\ 
 %%%%%%%%
 0 & 0 & \, 1 & 0\\ 
 %%%%%%%%
 0 & 0 & 0 & 1
\end{array}\right]
\, , %\cr&~~~
~ {(\gamma^1)}{}_a{}_b  = ( \sigma^3 
\otimes \sigma^3 ){}_a{}_b 
 = \left[\begin{array}{cccc} 
 1 & 0 & 0 & 0\\ 
 %%%%%%%%
 0 & -1 & 0 & 0\\ 
 %%%%%%%%
 0 & 0 & -1 & 0\\ 
 %%%%%%%%
0 & 0 & 0 & 1
\end{array}\right]
\, , \cr
& {(\gamma^2)}{}_a{}_b  = (\sigma^1 
\otimes  {\bf I}_2 ){}_a{}_b  
= \left[\begin{array}{cccc} 
 0 & 0 & 1 & 0\\ 
 %%%%%%%%
 0 & 0 & 0 & 1\\ 
 %%%%%%%%
 1 & 0 & 0 & 0\\ 
 %%%%%%%%
0 & 1 & 0 & 0
\end{array}\right]
\, , %\cr&~~~~~
~{(\gamma^3)}{}_a{}_b  = ( \sigma^3 
\otimes \sigma^1  ){}_a{}_b  
= \left[\begin{array}{cccc} 
 0 & 1 & 0 & 0\\ 
 %%%%%%%%
 1 & 0 & 0 & 0\\ 
 %%%%%%%%
 0 & 0 & 0 & -1\\ 
 %%%%%%%%
0 & 0 & -1 & 0
\end{array}\right]
~.
}    \label{KhyO7}
\ee
Use of these matrices then leads to the results
\begin{align*}
{\big \{} \,   D{}_{A} ~,~ D{}_{A} \, {\big\}} ~&=~  i  \, {\big[} \, (\g^{\mu
})_{1 \, 1}  \pa_{\mu} ~-~ (\g^{\mu})_{4 \, 4}  \pa_{\mu} \, {\big ]} ~-~ 
2 \, (\g^{\mu})_{1 \, 4}  \pa_{\mu}  ~~~, \cr
%%%%%%%%%%%%%%%%%%%%%%%%%%%%%
~&=~  i  \, {\big[} \, (\g^{0})_{1 \, 1}  \pa_{0} +  (\g^{1})_{1 \, 1}  
\pa_{1} ~-~ (\g^{0})_{4 \, 4}  \pa_{0} - (\g^{1})_{4 \, 4}  \pa_{1} \, {\big ]}  ~~~, 
\cr
~&=~ 0  \cr
&{~}  \cr
%%%%%%%%%%%%%%%%%%%%%%%%%%%%%
%%%%%%%%%%%%%%%%%%%%%%%%%%%%%
{\big \{} \,   D{}_{A} ~,~ D{}_{B} \, {\big\}} ~&=~   i  \, {\big[} \, 
(\g^{\mu})_{1 \, 2}  \pa_{\mu} ~+~ (\g^{\mu})_{3 \, 4}  \pa_{\mu} 
\, {\big ]} ~-~  \, (\g^{\mu})_{2 \, 4}  \pa_{\mu} ~+~  \, (\g^{\mu})_{1 \, 3}  
\pa_{\mu}  ~~~, \cr
%%%%%%%%%%%%%%%%%%%%%%%%%%%%%
~&=~   i  \, {\big[} \, (\g^{3})_{1 \, 2}  \pa_{3} ~+~ (\g^{3})_{3 \, 4}  
\pa_{3} \, {\big ]} ~-~  \, (\g^{2})_{2 \, 4}  
\pa_{2} ~+~  \, (\g^{2})_{1 \, 3}  
\pa_{2}  ~~~, \cr
~&=~ 0  \cr
&{~}  \cr
%%%%%%%%%%%%%%%%%%%%%%%%%%%%%
%%%%%%%%%%%%%%%%%%%%%%%%%%%%%
{\big \{} \,   D{}_{B} ~,~ D{}_{B} \, {\big\}} ~&=~ 
 i  \, {\big[} \, (\g^{\mu})_{2 \, 2}  \pa_{\mu} ~-~ (\g^{\mu})_{3 \, 3}  
\pa_{\mu} \, {\big ]} ~+~ 2 \, (\g^{\mu})_{2 \, 3}  
\pa_{\mu} ~~~, \cr
%%%%%%%%%%%%%%%%%%%%%%%%%%%%%
~&=~  i  \, {\big[} \, 
(\g^{0})_{2 \, 2}  \pa_{0} + (\g^{1})_{2 \, 2}  \pa_{1}
~-~ (\g^{0})_{3 \, 3}  \pa_{0} - (\g^{1})_{3 \, 3}  \pa_{1} 
\, {\big ]}  ~~~, \cr
~&=~ 0  \cr
&{~}  \cr
%%%%%%%%%%%%%%%%%%%%%%%%%%%%%
%%%%%%%%%%%%%%%%%%%%%%%%%%%%%
{\big \{} \,   D{}_{A} ~,~ {\Bar D}{}_{\dot B} \, {\big\}} ~&=~-\, i  \, 
{\big[} \, (\g^{3})_{1 \, 2}  \pa_{3} ~-~ (\g^{3})_{3 \, 4}  \pa_{3} \, 
{\big ]} ~-~ \, (\g^{2})_{2 \, 4}  \pa_{2} ~-~  \, (\g^{2})_{1 \, 3} 
 \pa_{2} ~~~, \cr
~&=~-\,   i  \, 2\,  {\big[} \,  \pa_{3} \,  ~-~ i\,  \pa_{2} 
 \, {\big ]} ~~~, \cr
&{~} 
\end{align*}
\begin{align}
%%%%%%%%%%%%%%%%%%%%%%%%%%%%%
%%%%%%%%%%%%%%%%%%%%%%%%%%%%%
{\big \{} \,   D{}_{A} ~,~ {\Bar D}{}_{\dot A} \, {\big\}} ~&=~-\, i  \, 
{\big[} \, (\g^{\mu})_{1 \, 1}  \pa_{\mu} ~+~ (\g^{\mu})_{4 \, 4}  
\pa_{\mu} \, {\big ]} ~~~, \nonumber\\*
~&=~-\, i  \, {\big[} \, (\g^{0})_{1 \, 1}  \pa_{0} + (\g^{1})_{1 \, 1}  
\pa_{1} ~+~ (\g^{0})_{4 \, 4}  \pa_{0} + (\g^{1})_{4 \, 4}  \pa_{1} 
\, {\big ]} ~~~,  \cr
~&=~-\, i  \, 2 \,  {\big[} \,  \pa_0 ~+~ \pa_1  \, {\big ]}  \cr
&{~}  \cr
%%%%%%%%%%%%%%%%%%%%%%%%%%%%%
%%%%%%%%%%%%%%%%%%%%%%%%%%%%%
{\big \{} \,   D{}_{B} ~,~ {\Bar D}{}_{\dot B} \, {\big\}} ~&=~ -\,  i  
\, {\big[} \, (\g^{\mu})_{2 \, 2}  \pa_{\mu} ~+~ (\g^{\mu})_{3 \, 3}  
\pa_{\mu} \,  {\big ]}  \cr
~&=~-\, i  \, {\big[} \, (\g^{0})_{2 \, 2}  \pa_{0} + (\g^{1})_{2 \, 2}  
\pa_{1} ~+~ (\g^{0})_{3 \, 3}  \pa_{0} + (\g^{1})_{3 \, 3}  \pa_{1} 
\, {\big ]} ~~~,   \cr
~&=~-\, i  \, 2 \, {\big[} \,  \pa_0 ~-~ \pa_1   \, {\big ]}  ~~~.
   \label{KhyO8}
\end{align}
If we define
\be {
\pa_{ A \, {\dot A}} ~=~ -\, 2 \,  {\big[} \,  \pa_0 ~+~ \pa_1  \, {\big ]} ~,~
\pa_{ A \, {\dot B}} ~=~ -\, 2  {\big[} \,  \pa_3 ~-~ i\, \pa_2   \, {\big ]} ~,~
\pa_{ B \, {\dot B}} ~=~ -\, 2  {\big[} \,  \pa_0 ~-~  \pa_1   \, {\big ]} 
}   \label{KhyO9}
\ee
then the operators $D{}_{A}$, $D{}_{B}$, $ {\Bar D}{}_{\dot A}$,
$ {\Bar D}{}_{\dot B}$, $\pa_{ A \, {\dot A}}$, $\pa_{ A \, {\dot B}}$,
and $\pa_{ B \, {\dot B}}$ satisfy the exact algebraic and hermiticity
properties of the corresponding objects defined in ``{\it {Superspace}},''
and we thus have an explicit dictionary. 

We define the $2 \times 2$ matrix $\partial_{\alpha\dot{\alpha}}$ as
\be
\eqalign{
\partial_{\alpha\dot{\alpha}} = \left(
\begin{array}{cc}
\partial_{A\dot{A}}  & \partial_{B\dot{B}} \\
\partial_{B\dot{A}} & \partial_{B\dot{B}} 
\end{array}
\right)
}     \label{KhyO10}
\ee
where $\partial_{B\dot{A}}$ is the complex conjugate of $\partial_{
A\dot{B}}$, i.e.
\be {
\partial_{B\dot{A}} \equiv \overline{\partial_{B\dot{A}}}  = -\, 2  {\big[} \,  
\pa_3 ~+~ i\, \pa_2   \, {\big ]}
}   \label{Khy11}
\ee
We have then explicitly
\be
\eqalign{
\partial_{\alpha\dot{\alpha}} =& -2\left(
\begin{array}{cc}
\pa_0 + \pa_1   & \pa_3 - i \pa_2  \\
\pa_3 + i \pa_2 & \pa_0 - \pa_1
\end{array}
\right) 
}    \label{Khy12}
\ee
Defining the soldering forms as
\be
\eqalign{
\tilde{\sigma}^0 =  \left(
\begin{array}{cc}
1   & 0 \\
0 & 1
\end{array}
\right)~~~,
&~~~\tilde{\sigma}^1 =  	
\left(
\begin{array}{cc}
1   & 0 \\
0 & -1
\end{array}
\right)~~~, \cr
\tilde{\sigma}^2 =
\left(	
\begin{array}{cc}
0  & -i \\
i & 0
\end{array}
\right)~~~,
&~~~\tilde{\sigma}^3 =  		
\left(
\begin{array}{cc}
0  & 1 \\
1 & 0
\end{array}
\right)~~~,
}    \label{Khy13}
\ee
we may neatly package $\partial_{\alpha\dot{\alpha}}$ as
\be
\eqalign{
\partial_{\alpha\dot{\alpha}} =& -2 \tilde{\sigma}^{\mu} \partial_\mu
}    \label{Khy14}
\ee
Finally, the two-component Weyl spinor operators denoted by $D_{\a}$
and ${\Bar D}{}_{\Dot \a}$ in {\em {Superspace}} \cite{SUSYBk} are given by
\be  \eqalign{
D_{\a} ~&=~ \left[
\begin{array}{c}
D_A   \\
D_B
\end{array}
\right]~~~,  ~~~
{\Bar D}{}_{\Dot \a} ~=~  \left[
\begin{array}{c}
{\Bar D}{}_{\dot A}  \\
{\Bar D}{}_{\dot B} 
\end{array}
\right]~~~.
 }    \label{Khy15}
\ee 

With this completed dictionary, we note that in the four dimensional $\cal N$
= 1 conventions of {\em {Superspace}}, one can define a quantum number
${\widehat \chi}{}_{\rm o}$ that appears in the following definition
\be  {
{\widehat \chi}{}_{\rm o} \, \square  ~=~ -\,  {\big [} \,\,  2 \, D{}^{\a} {\Bar D}{}^2   D{}_{\a} 
~+~ \square \,\,  {\big ]}
 }    \label{Khy16}
\ee 
and we wish to calculate the value of this quantum number on the spinor
component fields that appear in the chiral, vector and tensor superfields. 
It is appropriate here to note that in principle and on a general superfield
there may be no value ${\widehat \chi}{}_{\rm o}$ for which this equation
possesses a solution.  However, whenever a superfield is subject to a
sufficient number of spinorial differential constraints, this is not a concern.  
In particular, for superfields that represent {\em {irreducible}} supermultiplets, 
such constraints are enforced.  This is most certainly the case for the chiral
($\Phi$), vector ($W_{\a}$) and tensor supermultiplets ($G$).

We recall that these superfields can be described in the following manner
by use of the respective pre-potentials $U$, $V$, and $\U_{\a}$:
\newline \indent
(a.)
$\Phi$ = ${\Bar D}{}^2 \, U$, where $U \, \ne \, {\Bar U}$,
\newline \indent
(b.) 
$W_{\a}$ = $i \, {\Bar D}{}^2 \, D{}_{\a} \, V$  where $V \, = \, {\Bar V}$, and
\newline \indent
(c.) 
$G$ = $ D{}^{\a} \,  {\Bar D}{}^2 \, \U_{\a}\,+\, {\rm {h.\, c.}}$ 
\newline \noindent
which will be permit a rapid determination of the value of ${\widehat \chi}{}_{\rm o}$
on the spinor component in each supermultiplet.  These spinor components
are given respectively by
\be {
\psi_{\a} ~\equiv ~  D{}_{\a} \, \Phi \, {\big |}  ~~,~~ \lambda_{\a} 
~\equiv ~  W{}_{\a} \,  {\big |}  ~~,~~  \chi{}_{\a} ~\equiv ~  D{}_{\a} \, G \, 
{\big |}  ~~,~~
 }    \label{Khy17}
\ee 
which leads us to three calculations:
\begin{align*}
{\widehat \chi}{}_{\rm o} \, \square \left( D{}_{\b} \, \Phi \, {\big |} \right)
 ~&=~ -\,  {\big [} \,\,  2 \, D{}^{\a} {\Bar D}{}^2   D{}_{\a} 
~+~ \square \,\,  {\big ]} \, \left( D{}_{\b} \, \Phi \, {\big |} \right)    \cr
 ~&=~  {\big [} \,\,  -\, 2 \, D{}^{\a} {\Bar D}{}^2   D{}_{\a} \left( D{}_{\b} \, \Phi \, {\big |} \right)
~-~ \square\, \left( D{}_{\b} \, \Phi \, {\big |} \right) \,\,  {\big ]}    \cr
 ~&=~  {\big [} \,\,   2 \, D{}_{\b} {\Bar D}{}^2  \left(   D^2 \, \Phi \, {\big |} \right)
~-~ \square\, \left( D{}_{\b} \, \Phi \, {\big |} \right) \,\,  {\big ]}    \cr
 ~&=~ \square\, \left( D{}_{\b} \, \Phi \, {\big |} \right)  ~~
 ~~~~~~~~~~~~~~~~~~~~~~~~~~~~~~~~~~~~~~~\to~ {\widehat \chi}{}_{\rm o} ~=~
 + \, 1  ~~~,  
 \end{align*}
 \begin{align}
{\widehat \chi}{}_{\rm o} \, \square \left( W{}_{\b} \, {\big |} \right)
 ~&=~ -\,  {\big [} \,\,  2 \, D{}^{\a} {\Bar D}{}^2   D{}_{\a} 
~+~ \square \,\,  {\big ]} \,  \left( W{}_{\b} \, {\big |} \right)  \nonumber\\*
 ~&=~  {\big [} \,\,  -\, 2 \, D{}^{\a} {\Bar D}{}^2   D{}_{\a}  \left( W{}_{\b} \, {\big |} \right)
~-~ \square\,  \left( W{}_{\b} \, {\big |} \right) \,\,  {\big ]}    \cr
 ~&=~  {\big [} \,-\,  i 2 \, \left( \, D{}^{\a} {\Bar D}{}^2 D{}_{\a}  \, {\Bar D}{}^2 \, D{}_{\b} \, V \right)
~-~ \square\,  \left( W{}_{\b} \, {\big |} \right) \,\,  {\big ]}    \cr
 ~&=~-\,  \square\,  \left( W{}_{\b} \, {\big |} \right)  ~~~
 ~~~~~~~~~~~~~~~~~~~~~~~~~~~~~~~~~~~~\to~ {\widehat \chi}{}_{\rm o} ~=~
 - \, 1  ~~~,  \cr
{\widehat \chi}{}_{\rm o} \, \square \left( D{}_{\b} \, G \, {\big |} \right)
 ~&=~ -\,  {\big [} \,\,  2 \, D{}^{\a} {\Bar D}{}^2   D{}_{\a} 
~+~ \square \,\,  {\big ]} \, \left( D{}_{\b} \, G \, {\big |} \right)    \cr
 ~&=~  {\big [} \,\,  -\, 2 \, D{}^{\a} {\Bar D}{}^2   D{}_{\a} \left( D{}_{\b} \, G \, {\big |} \right)
~-~ \square\, \left( D{}_{\b} \, G \, {\big |} \right) \,\,  {\big ]}    \cr
 ~&=~  {\big [} \,\,  -\,  2 \, D{}^{\a} {\Bar D}{}^2 D{}_{\a} \, D{}_{\b}
  \left(  D{}^{\g} \,  {\Bar D}{}^2 \, \U_{\g}\,+\, {\rm {h.\, c.}} \, {\big |} \right)
~-~ \square\, \left( D{}_{\b} \,G \, {\big |} \right) \,\,  {\big ]}    \cr
~&=~  {\big [} \,\,    2 \, D{}_{\b} {\Bar D}{}^2  D{}^2 \left(D{}^{\g} \,  {\Bar 
D}{}^2 \, \U_{\g}\,+\, {\rm {h.\, c.}} \, {\big |} \right) ~-~ \square\, \left( D
{}_{\b} \,G \, {\big |} \right) \,\,  {\big ]}    \cr
 ~&=~ -\,  \square\, \left( D{}_{\b} \, G \, {\big |} \right)  ~~~
 ~~~~~~~~~~~~~~~~~~~~~~~~~~~~~~~~~~\to~ {\widehat \chi}{}_{\rm o} ~=~
 - \, 1  ~~~.
    \label{Khy18}
\end{align}
where respectively we have used the identities, 
\be {
 {\Bar D}{}^2    D^2 \, \Phi  ~=~ \square \Phi ~~,~~ 
  D{}^{\a} {\Bar D}{}^2 D{}_{\a}  \, {\Bar D}{}^2 ~=~ 0  ~~,~~
 {D}{}^2 D{}^{\g}   ~=~ 0 ~~~.
 }    \label{Khy19}
\ee 

These calculations beautifully demonstrate the result that $\chi{}_{\rm o}$ =
${\widehat \chi}{}_{\rm o}$ on the three respective valise adinkras on one
side of the calculation and the three respective supermultiplets on the other.  
In other words, this is another example of SUSY holography at work.

The result of this section shows that the valise adinkra-based calculation
(\ref{Chr0M3a}) leads to the same result as the 4D, $\cal N$ = 1 superfield 
calculation of the operator defined in (\ref{Khy16}).  In other words, the
information in operator in  (\ref{Khy16}) is the same as the information in
(\ref{Chr0M3a}).  Thus, for some operators acting on 4D, $\cal N$ = 1
superfields, equivalent operators can be found to act on valise adinkras.
This opens up the possibility that there may be other such operators
for which this statement holds.

However, the real power of the valise adinkra viewpoint in these examples
has been to easily identify the quantum number ${\widehat \chi}{}_{\rm o}$ 
that exists among 4D, $\cal N$ = 1 superfields that determines when these
form an off-shell representation and to explain `why' ${\widehat \chi}{}_{\rm 
o}$ must a priori vanish when summed over 4D, $\cal N$ = 1 superfields
to construct 4D, $\cal N$ = 2 superfields.

%%%%%%%%%%%%%%%%%%%%%%%%%%%%%%%%%%%%%%%%%%
 %%%%%%%%%%%%%%%%%%%%%%%%%%%%%%%%%%%%%%%%%%
 %%%%%%%%%%%%%%%%%%%%%%%%%%%%%%%%%%%%%%%%%%
\section{A Garden Algebra/Unconstrained Superspace Prepotential
Formulation No-Go Conjecture}

$~~~~$ The results of chapter five also provide the basis for making a
conjecture about the relation of representations of ${ { {\cal GR} ({\rm d}, \,  
{N})}}$, representations of ${ { {\cal GR} ({\rm d}_L, \, {\rm d}_R, \,  { 
N})}}$, and unconstrained prepotential formulations of higher dimensional 
supermultiplets.  As can be seen from the models studied earlier in this
paper, whenever a supermultiplet is a representation of the ${ { {\cal GR} 
({\rm d}, \,  {N})}}$ algebras, it {\em {does}} {\em {not}} possess an off-shell 
central charge.  Alternately, whenever a supermultiplet is a representation 
of the ${ { {\cal GR} ({\rm d}_L, \, {\rm d}_R, \,  {N})}}$ algebras,  it {\em {does}} 
possess an off-shell central charge.

There is another observation about supermultiplets that is interesting to
note in the context of unconstrained Salam-Strathdee superfields.  An
 unconstrained Salam-Strathdee superfield is one that is not subject to
 any type of differential (either spacetime nor D-operator) constraint.
All superfields that are quantizable, can be expressed in terms of
 unconstrained Salam-Strathdee superfields.  
 
 There is a direct relation between the component field formulation of a
 supermultiplet that does not possess off-shell central charges and
 their expression in terms of  unconstrained Salam-Strathdee superfields.
 The component fields of a supermultiplet come in different engineering
 dimensions.  In the adinkra represents, this assignment of engineering
 dimension corresponds to the height at which a node associated with
 a component field appears in the adinkra.  
 
 When one identifies the highest fields in the adinkra representing 
 a supermultiplet with no off-shell central charge, one has identified
 the  unconstrained Salam-Strathdee superfields that describes the
 supermultiplet.
 
 This brings us to a conjecture:

${~~~~~}$ {\it {Only supermultiplets that do {\underline 
{not}} contain off-shell central charges \newline ${~~~~~~~~~}$
are representations of ${ { {\cal GR} ({\rm d}, \,  
{N})}}$ algebras that can be described by
\newline ${~~~~~~~~~}$ unconstrained Salam-Strathdee superfields uniquely determined
by the
\newline ${~~~~~~~~~}$ 
 highest engineering dimension component fields with no spacetime 
\newline ${~~~~~~~~~}$ 
 derivatives.}} 

 %%%%%%%%%%%%%%%%%%%%%%%%%%%%%%%%%%%%%%%%%%
 %%%%%%%%%%%%%%%%%%%%%%%%%%%%%%%%%%%%%%%%%%
 %%%%%%%%%%%%%%%%%%%%%%%%%%%%%%%%%%%%%%%%%%
\section{Conclusion}

$~~~~$ We hope to have convinced the reader that our efforts have uncovered
a \emph{new} quantum number ($\chi_{\rm o}$) in supersymmetrical field theory.  The
value of this quantum number for some familiar 4D, $\cal N$ $=$ 1 supermultiplets
is shown in the table below.

\begin{table}[!h]
\centering
\caption{The new quantum number $\chi_0$ for the 4D, ${\mathcal N}=1$ chiral (CM), vector (VM), tensor (TM), real scalar (RSS), complex linear (CLM), conformal supergravity (cSG), old-minimal supergravity (mSG), and non-minimal supergravity ($\not$mSG) multiplets \cite{G-1,GII-III}.}
\begin{tabularx}{\linewidth}{c|Z|Z|Z|Z|Z|Z|Z|c|}
	& CM
	& VM
		& TM
	& RSS
	&CLS & cSG & mSG & $\not$mSG
	\\
\hline
$\chi_0$	&$1$
	&$-1$
		& $-1$ 	 	
	& $0$
	& $-1$	 
	& $-2$
	& $-1$ 
	& $-3$\\
	\hline
\end{tabularx}
\label{t:tc1}
\end{table}

This table implies that there are two distinct ways to construct Dirac
particles in supersymmetrical theories.  A standard approach to embedding
Dirac particles into  4D, $\cal N$ $=$ 1 models is to use a pair of chiral
superfields that may be denoted by $\Phi_+$ and $\Phi_-$ corresponding
to a $\chi_{\rm o}$ $=$ 2 system.  In a number of our past works \cite{CNM},
\cite{CNMmassless}, and \cite{ProjectiveSuperspaceGK}, it has been
advocated that an alternate approach to embedding Dirac particles into  
4D, $\cal N$ $=$ 1 models is to use a `CMN pair' consisting of one chiral
superfield $\Phi$ and one complex linear superfield $\S$ corresponding
to a $\chi_{\rm o}$ $=$ 0 system.  

One of the amusing analogies to note is that with respect to off-shell 4D, $\cal N$
= 2 supersymmetry, the adinkra quantum number $\chi{}_{\rm o}$ defined
on 4D, $\cal N$ = 1 supermultiplets acts just like color in hadronic physics!
It seems likely that off-shell 4D, $\cal N$ = 2 supersymmetry representations
most have vanishing adinkra quantum number $\chi{}_{\rm o}$ just as
baryons must have vanishing color. 

Our present work shows that with regard
to 4D, $\cal N$ $=$ 2 SUSY this new quantum number matters.  As far as
we can tell, all Dirac fermions in off-shell  4D, $\cal N$ $=$ 2 systems
have  $\chi_{\rm o}$ $=$ 0.  This raises numbers of questions.
Does this have implications for 4D, $\cal N$ $=$ 1 SUSY systems, including
phenomenology?  It is known that there exist 4D, $\cal N$ $=$ 1
duality transformations between  $\chi_{\rm o}$ $=$ 0 and
 $\chi_{\rm o}$ $=$ 2 systems.  Do our results imply that no
 such 4D, $\cal N$ $=$ 2 duality transformations exist?  Needless
 to say all of this is very strange and `funny.'

 \vspace{.05in}
 \begin{center}
 \parbox{4in}{{\it ``The most exciting phrase to hear in
 science, the 
\newline $~~\,$ one that heralds new discoveries, is
 not `Eureka!' 
 \newline $~~\,$ but
 `That's funny...' ''}\,\,-\,\, Isaac Asimov}
 \end{center}
 
\noindent
{\bf Acknowledgements}\\[.1in] \indent
This work was partially supported by the National Science Foundation grants 
PHY-0652983 and PHY-0354401. This research was also supported in part by the 
University of Maryland Center for String \& Particle Theory.    

\noindent
{\bf Dedication}\\[.1in] \indent
This work is dedicated to our valued colleague O.\ W.\ Greenberg and his many
years of contributions to the field.  It is fitting we believe that our discovery of a
color-like quantum number with regard to off-shell supersymmetry should occur
contemporaneously with the celebration of his groundbreaking work of 1964 leading
to the color concept now widely accepted in hadronic physics as the basis for
QCD.

\newpage
\noindent
{\Large\bf Appendix A: Chiral $+$ Chiral L, R Matrices}

In the following series of appendices, we give the explicit forms of the L-matrices and
R-matrices discussed in the third chapter.  We use the compact `Boolean Factor/Cycle'
notation introduced in our work of \cite{permutadnk}.   The explicit form of the 8 $\times$ 
8 L-matrices and R-matrices that appear in (\ref{eq:CCLR}) are found to be:
$$ \eqalign{
\left( {\rm L}{}_{1}\right)  ~=~ \left[\begin{array}{cc}
(10)_b   (243) &  0  \\
0 &   (10)_b    (243)  \\
\end{array}\right]   ~~&,~~
\left( {\rm L}{}_{2}\right)  ~=~ \left[\begin{array}{cc}
(12)_b   (123) &  0  \\
0 &   (12)_b    (123)  \\
\end{array}\right]  ~~,
\cr
\left( {\rm L}{}_{3}\right)  ~=~ \left[\begin{array}{cc}
(6)_b   (134) &  0  \\
0 &   (6)_b    (134)  \\
\end{array}\right]   ~~~~~&,~~
\left( {\rm L}{}_{4}\right)  ~=~ \left[\begin{array}{cc}
(0)_b   (142) &  0  \\
0 &   (0)_b    (142)  \\
\end{array}\right]  ~~~~~,  \cr
 \left( {\rm L}{}_{5}\right)  ~=~ \left[\begin{array}{cc}
0 &  (15)_b   (243)  \\
(0)_b   (243) &   0  \\
\end{array}\right]   ~~&,~~
\left( {\rm L}{}_{6}\right)  ~=~ \left[\begin{array}{cc}
0  &  (9)_b   (123)  \\
(6)_b    (123) &   0  \\
\end{array}\right]  ~~~~~, \cr
\left( {\rm L}{}_{7}\right)  ~=~ \left[\begin{array}{cc}
0 &  (3)_b   (134)  \\
(12)_b   (134) &   0  \\
\end{array}\right]   ~~&,~~
\left( {\rm L}{}_{8}\right)  ~=~ \left[\begin{array}{cc}
0  &  (5)_b   (142)  \\
(10)_b    (142) &   0  \\
\end{array}\right]  ~~~. }
$$
$$~~$$
$$~~$$

\noindent
{\Large\bf Appendix B: Chiral $+$ Vector L, R Matrices}

The explicit form of the 8 $\times$ 8 L-matrices and R-matrices derived from the
case of case of the chiral $+$ vector supermultiplets and that are analogous to those that
appear in (\ref{eq:CCLR}) are found to
be:
$$ \eqalign{
\left( {\rm L}{}_{1}\right)  ~=~ \left[\begin{array}{cc}
(10)_b   (243) &  0  \\
0 &   (10)_b    (1243)  \\
\end{array}\right]   ~~&,~~
\left( {\rm L}{}_{2}\right)  ~=~ \left[\begin{array}{cc}
(12)_b   (123) &  0  \\
0 &   (12)_b    (23)  \\
\end{array}\right] ~~~~~, \cr
\left( {\rm L}{}_{3}\right)  ~=~ \left[\begin{array}{cc}
(6)_b   (134) &  0  \\
0 &   (0)_b    (14)  \\
\end{array}\right]   ~~~~~~~~&,~~
\left( {\rm L}{}_{4}\right)  ~=~ \left[\begin{array}{cc}
(0)_b   (142) &  0  \\
0 &   (6)_b    (1342)  \\
\end{array}\right]  ~~~~~,   \cr
 \left( {\rm L}{}_{5}\right)  ~=~ \left[\begin{array}{cc}
0 &  (2)_b   (243)  \\
(13)_b   (1243) &   0  \\
\end{array}\right]   ~~~~&,~~
\left( {\rm L}{}_{6}\right)  ~=~ \left[\begin{array}{cc}
0  &  (4)_b   (123)  \\
(11)_b    (23) &   0  \\
\end{array}\right]  ~~~~~~,  \cr
\left( {\rm L}{}_{7}\right)  ~=~ \left[\begin{array}{cc}
0 &  (14)_b   (134)  \\
(7)_b   (14) &   0  \\
\end{array}\right]   ~~~~~~\,~&,~~
\left( {\rm L}{}_{8}\right)  ~=~ \left[\begin{array}{cc}
0  &  (8)_b   (142)  \\
(1)_b    (1342) &   0  \\
\end{array}\right]  ~~\,~~,
} $$
\newpage

\noindent
{\Large\bf Appendix C: Chiral $+$ Tensor L, R Matrices}

The explicit form of the 8 $\times$ 8 L-matrices and R-matrices derived 
from the case of case of the chiral $+$ tensor supermultiplets and that 
are analogous to those that appear in (\ref{eq:CCLR}) are found to
be:
$$ \eqalign{
\left( {\rm L}{}_{1}\right)  ~=~ \left[\begin{array}{cc}
(10)_b   (243) &  0  \\
0 &   (14)_b    (234)  \\
\end{array}\right]   ~~&,~~
\left( {\rm L}{}_{2}\right)  ~=~ \left[\begin{array}{cc}
(12)_b   (123) &  0  \\
0 &   (4)_b    (124)  \\
\end{array}\right] ~~~~~, \cr
\left( {\rm L}{}_{3}\right)  ~=~ \left[\begin{array}{cc}
(6)_b   (134) &  0  \\
0 &   (8)_b    (132)  \\
\end{array}\right]   ~~~~~&,~~
\left( {\rm L}{}_{4}\right)  ~=~ \left[\begin{array}{cc}
(0)_b   (142) &  0  \\
0 &   (2)_b    (143)  \\
\end{array}\right]  ~~~~\,~~,  \cr
 \left( {\rm L}{}_{5}\right)  ~=~ \left[\begin{array}{cc}
0 &  (11)_b   (243)  \\
(0)_b   (234) &   0  \\
\end{array}\right]   ~~\,~&,~~
\left( {\rm L}{}_{6}\right)  ~=~ \left[\begin{array}{cc}
0  &  (13)_b   (123)  \\
(10)_b    (124) &   0  \\
\end{array}\right]  ~~~~,   \cr
\left( {\rm L}{}_{7}\right)  ~=~ \left[\begin{array}{cc}
0 &  (7)_b   (134)  \\
(6)_b   (132) &   0  \\
\end{array}\right]   ~~~~\,~&,~~
\left( {\rm L}{}_{8}\right)  ~=~ \left[\begin{array}{cc}
0  &  (1)_b   (142)  \\
(12)_b    (143) &   0  \\
\end{array}\right]  ~~~~~,
} $$
$$~~$$
$$~~$$
\noindent
{\Large\bf Appendix D: Vector $+$ Vector L, R Matrices }

The explicit form of the 8 $\times$ 8 L-matrices and R-matrices derived from the
 case of the vector $+$ vector supermultiplets and that are analogous to 
those that appear in (\ref{eq:CCLR}) are found to be
$$ \eqalign{
\left( {\rm L}{}_{1}\right)  ~&=~ \left[\begin{array}{cc}
b_+ (10)_b    (1243) &  0  \\
0 &  a_+ (10)_b    (1243)  \\
\end{array}\right]   ~~,~~  \cr
\left( {\rm L}{}_{2}\right)  ~&=~ \left[\begin{array}{cc}
b_+ (12)_b    (23) &  0  \\
0 & a_+  (12)_b    (23)  \\
\end{array}\right] ~~~~~~~~, \cr
\left( {\rm L}{}_{3}\right)  ~&=~ \left[\begin{array}{cc}
b_+ (0)_b    (14) &  0  \\
0 &  a_+ (0)_b    (14)  \\
\end{array}\right]   ~~~~~~~~~~~,~~  \cr
\left( {\rm L}{}_{4}\right)  ~&=~ \left[\begin{array}{cc}
b_+ (6)_b    (1342) &  0  \\
0 &  a_+ (6)_b    (1342)  \\
\end{array}\right]  ~~~~~,   \cr
 \left( {\rm L}{}_{5}\right)  ~&=~ \left[\begin{array}{cc}
 0 &  b_- (10)_b   (1243)  \\
a_- (10)_b   (1243) &   0  \\
\end{array}\right]   ~~,~~ \cr
\left( {\rm L}{}_{6}\right)  ~&=~ \left[\begin{array}{cc}
0  & b_- (12)_b    (23)  \\
a_- (12)_b    (23) &   0  \\
\end{array}\right]  ~~~~~~\,~~\,~~,  }$$
$$ \eqalign{
\left( {\rm L}{}_{7}\right)  ~&=~ \left[\begin{array}{cc}
0 & b_- (0)_b   (14)  \\
a_- (0)_b   (14) &   0  \\
\end{array}\right]   ~~~~~~~~~~,~~  \cr
\left( {\rm L}{}_{8}\right)  ~&=~ \left[\begin{array}{cc}
0  & b_- (6)_b    (1342)  \\
a_- (6)_b    (1342) &   0  \\
\end{array}\right]  ~~\,~~,
} $$
where
$$
a_{\pm} = \cos \left( \frac{m\pi}{2}\right) \pm \sin \left( \frac{m\pi}{2}\right) 
~~~,~~~b_{\pm} = 
\cos \left( \frac{n\pi}{2}\right) \pm \sin \left( \frac{n\pi}{2}\right) ~~~.
$$

$$~$$
\noindent
{\Large\bf Appendix E: Tensor $+$ Tensor L, R Matrices}

The explicit form of the 8 $\times$ 8 L-matrices and R-matrices derived from the
case of the tensor $+$ tensor supermultiplets and that are analogous to 
those that appear in (\ref{eq:CCLR}) are found to be
$$ \eqalign{
\left( {\rm L}{}_{1}\right)  ~&=~ \left[\begin{array}{cc}
b_+ (14)_b    (234) &  0  \\
0 &  a_+ (14)_b    (234)  \\
\end{array}\right]   ~~,~~  \cr
\left( {\rm L}{}_{2}\right)  ~&=~ \left[\begin{array}{cc}
b_+ (4)_b    (124) &  0  \\
0 & a_+  (4)_b    (124)  \\
\end{array}\right] ~~~~~~~~, \cr
\left( {\rm L}{}_{3}\right)  ~&=~ \left[\begin{array}{cc}
b_+ (8)_b    (132) &  0  \\
0 &  a_+ (8)_b    (132)  \\
\end{array}\right]   ~~~~~~~~~~~,~~  \cr
\left( {\rm L}{}_{4}\right)  ~&=~ \left[\begin{array}{cc}
b_+ (2)_b    (143) &  0  \\
0 &  a_+ (2)_b    (143)  \\
\end{array}\right]  ~~~~~,   \cr
 \left( {\rm L}{}_{5}\right)  ~&=~ \left[\begin{array}{cc}
 0 &  b_- (14)_b   (234)  \\
a_- (14)_b   (234) &   0  \\
\end{array}\right]   ~~,~~ \cr
\left( {\rm L}{}_{6}\right)  ~&=~ \left[\begin{array}{cc}
0  & b_- (4)_b    (124)  \\
a_- (4)_b    (124) &   0  \\
\end{array}\right]  ~~~~~~\,~~\,~~,  }$$
$$ \eqalign{
\left( {\rm L}{}_{7}\right)  ~&=~ \left[\begin{array}{cc}
0 & b_- (8)_b   (132)  \\
a_- (8)_b   (132) &   0  \\
\end{array}\right]   ~~~~~~~~~~,~~  \cr
\left( {\rm L}{}_{8}\right)  ~&=~ \left[\begin{array}{cc}
0  & b_- (2)_b    (143)  \\
a_- (2)_b    (143) &   0  \\
\end{array}\right]  ~~\,~~,
} $$

\noindent
{\Large\bf Appendix F: Vector $+$ Tensor L, R Matrices}

The explicit form of the 8 $\times$ 8 L-matrices and R-matrices derived from the
case of the vector $+$ tensor supermultiplets and that are analogous to 
those that appear in (\ref{eq:CCLR}) are found to be
$$ \eqalign{
\left( {\rm L}{}_{1}\right)  ~&=~ \left[\begin{array}{cc}
{b}_+ (14)_b    (234)&  0  \\
0 & {a}_+ (10)_b    (1243)    \\
\end{array}\right]   ~~~~,~~  \cr
\left( {\rm L}{}_{2}\right)  ~&=~ \left[\begin{array}{cc}
 {b}_+   (4)_b    (124) &  0  \\
0 & {a}_+ (12)_b    (23)   \\
\end{array}\right] ~~~~~~~~, \cr
\left( {\rm L}{}_{3}\right)  ~&=~ \left[\begin{array}{cc}
{b}_+ (8)_b    (132)  &  0  \\
0 & {a}_+ (0)_b    (14)  \\
\end{array}\right]   ~~~~~\,~~~~,~~  \cr
\left( {\rm L}{}_{4}\right)  ~&=~ \left[\begin{array}{cc}
 {b}_+ (2)_b    (143) &  0  \\
0 & {a}_+ (6)_b    (1342)  \\
\end{array}\right]  ~~~~~~~,   \cr
 \left( {\rm L}{}_{5}\right)  ~&=~ \left[\begin{array}{cc}
 0 &{b}_- (14)_b    (234)  \\
 {a}_- (10)_b   (1243)  &   0  \\
\end{array}\right]   ~~~~,~~ \cr
\left( {\rm L}{}_{6}\right)  ~&=~ \left[\begin{array}{cc}
0  & {b}_-  (4)_b    (124) \\
{a}_- (12)_b    (23)  &   0  \\
\end{array}\right]  ~~~~~~\,~~,  \cr
\left( {\rm L}{}_{7}\right)  ~&=~ \left[\begin{array}{cc}
0 & {b}_- (8)_b    (132)  \\
{a}_- (0)_b   (14) &   0  \\
\end{array}\right]   ~~~~~~~~~,~~  \cr
\left( {\rm L}{}_{8}\right)  ~&=~ \left[\begin{array}{cc}
0  & {b}_-  (2)_b    (143) \\
 {a}_- (6)_b    (1342) &   0  \\
\end{array}\right]  ~~~~~~.
} $$

\noindent
{\Large\bf Appendix G: $q{\cal {GR}}$ Bracket Example Calculations}

In this appendix, we will simply demonstrate two example of how the 
$q{\cal {GR}}$ bracket defined in chapter two can be used.  In the first
case we show it leads to a very different perspective using the usual 
Pauli matrices.

We, of course, use their conventional definitions
$$
\s^1 ~=~  \left(\begin{array}{cc}
0 &  1  \\
1 &  0  \\
\end{array}\right)  ~~,~~ 
\s^2 ~=~  \left(\begin{array}{cc}
0 &  -i  \\
i &  0   \\
\end{array}\right)  ~~,~~ 
\s^3 ~=~  \left(\begin{array}{cc}
1 &  0  \\
0 &  -1   \\
\end{array}\right)  ~~,~~ 
\eqno(H.1)
$$
implying
$$
\s^i \, \s^j ~=~ \delta^{i \, j} \, {\bm {\rm I}}~+~ i \, \epsilon^{i \, j \, k} \, \s^k
~~\to~~ 
\left[ \, \s^i ~,~ \s^j \, \right] ~=~ i 2 \, \epsilon^{i \, j \, k} \, \s^k  ~~~,
\eqno(H.2) 
$$
the usual commutator algebra.  We note that
$$
\left(\s^1 \right) {}^t~=~ + \,  \left(\s^1 \right) ~~,~~ 
\left(\s^2 \right) {}^t~=~ - \,  \left(\s^2 \right) ~~,~~ 
\left(\s^3 \right) {}^t~=~ + \,  \left(\s^3 \right) ~~.
\eqno(H.3) $$
Under the action of the $q{\cal {GR}}$ brackets for $q$ =  1, we have
$$
{ \big[  } \,  \s^1 ~,~ \s^2 \, {\big ]}_{(1){\cal {GR}}>}  ~=~ 0
~~,~~ { \big[  } \,  \s^2 ~,~ \s^3 \, {\big ]}_{(1){\cal {GR}}>}  ~=~ 0
~~,~~ { \big[  } \,  \s^3 ~,~ \s^1 \, {\big ]}_{(1){\cal {GR}}>}  ~=~ 
 i 2 \,  \, \s^2 ~~,
\eqno(H.4)$$
$$
{ \big[  } \,  \s^1 ~,~ \s^2 \, {\big ]}_{(1){\cal {GR}}<}  ~=~ 0
~~,~~ { \big[  } \,  \s^2 ~,~ \s^3 \, {\big ]}_{(1){\cal {GR}}<}  ~=~ 0
~~,~~ { \big[  } \,  \s^3 ~,~ \s^1 \, {\big ]}_{(1){\cal {GR}}<}  ~=~ 
 i 2 \,  \, \s^2 ~~.
\eqno(H.5)$$
The results in (H.4) and the ones in (H.5) each separately result imply that a 
Jacobi-like condition is satisfied by the $q{\cal {GR}}$ bracket for $q$ 
=  1 and the Pauli matrices. So a structure not dissimilar to a Lie algebra 
emerges.  Since the Pauli matrices can be identified as the generators 
of the $su(2)$ algebra, replacing them by the generators for $su(3)$ 
leads to more interesting results.  It might be of interest to investigate
whether such a replacement also lead to a structure not dissimilar to a 
Lie algebra.

\end{document}